\documentclass{WileyMSP-template}

\usepackage{url}
\usepackage{graphicx}
\usepackage{float}
\usepackage{stfloats}
\usepackage{textcomp}
\usepackage{xcolor}
\usepackage{lipsum}
\usepackage{rotating}
\usepackage{tikz}
\usepackage{afterpage}
\usepackage{verbatim}
\usepackage{caption}
\usepackage{siunitx}
\usepackage{ragged2e}
\usepackage[export]{adjustbox}

\usepackage[export]{adjustbox}
\usepackage{ragged2e}
\usepackage[export]{adjustbox}
\usepackage{comment}
\usepackage[utf8]{inputenc}
\usepackage[english]{babel}
\usepackage[
backend=biber,
sorting=none
]{biblatex}
\usepackage{longtable}
\usepackage{array}
\usepackage{booktabs}
\usepackage{subcaption}
\usepackage{amsmath}
\usepackage[margins]{trackchanges}
\usepackage{float}
\usepackage{placeins}
\usepackage{multirow}
\begin{document}
%\pagestyle{fancy}
%\rhead{\includegraphics[width=2.5cm]{vch-logo.png}}
\title{Organic log-domain integrator synapse}

\maketitle

% Author: Please give full first and last names for authors and include * after the name of all corresponding authors

\author{Mohammad Javad Mirshojaeian Hosseini,}
%\author{Author Two}
\author{Elisa Donati,}
\author{Giacomo Indiveri,}
\author{Robert A. Nawrocki*}

% Dedication

%\dedication{Optional dedication here. If no dedication is required, please leave blank}

% Affiliations: Please provide adacemic titles (Prof. or Dr.) for all authors where applicable, and include an institutional email address for all corresponding authors
\begin{affiliations}
M. J. Mirshojaeian Hosseini, R. A. Nawrocki\\
School of Engineering Technology, Purdue University, West Lafayette, IN, United States of America\\
Email Address: mmirshoj@purdue.edu, robertnawrocki@purdue.edu\\
\hfill \break

E. Donati, G. Indiveri\\
Institute of Neuroinformatics, University of Zurich and ETH Zurich, Zurich, Switzerland\\
Email Address: elisa@ini.uzh.ch, giacomo@ini.uzh.ch

\end{affiliations}
% Keywords: Please provide a minimum of three and a maximum of seven keywords, separated by commas

\keywords{Neuromorphics, Organic Neuromorphics, Flexible organic synaptic circuit, Log-domain integrator synapse, Biologically plausible time constant}

% Abstract should be written in the present tense and impersonal style (i.e., avoid we), and be at most 200 words long

\begin{abstract}
  Synapses play a critical role in memory, learning, and cognition. Their main functions include converting pre-synaptic voltage spikes to post-synaptic currents, as well as scaling the input signal. Several brain-inspired architectures have been proposed to emulate the behavior of biological synapses.
  While these are useful to explore the properties of nervous systems, the challenge of making biocompatible and flexible circuits with biologically plausible time constants and tunable gain remains.
  Here, a physically flexible organic log-domain integrator synaptic circuit is shown to address this challenge.
In particular, the circuit is fabricated using organic-based materials that are electrically active, offer flexibility and biocompatibility, as well as time constants (critical in learning neural codes and encoding spatiotemporal patterns) that are biologically plausible.
Using a 10 \,nF synaptic capacitor, the time constant reached 126\,ms and 221\,ms before and during bending, respectively.
The flexible synaptic circuit is characterized before and during bending, followed with studies on the effects of weighting voltage, synaptic capacitance, and disparity in pre-synaptic signals on the time constant.
\end{abstract}

% Text: Please use section headings and subheadings as specified below. For communications, all section headings apart from Experimental Section should be removed
% Please make the first reference to a display item bold: \textbf{Figure 1}
% Do not abbreviate Figure, Equation, etc.; display items are always singular, i.e., Figure 1 and 2.
% Equations are always singular, i.e., Equation 1 and 2, and should be inserted using the {equation} environment, not as graphics
% Please do not use footnotes in the text, additional information can be added to the Reference list.

\section{Introduction}
Since the early 1990s disadvantages of conventional von Neumann computing architectures, including time-multiplexed serial processing, explicit programming, and high power consumption, have driven the  development of biologically inspired electronic circuits to emulate sensory processing systems and spiking neural networks, known as Neuromorphic Engineering~\cite{Mead20,b2,b1}.
Neuromorphic architectures are characterized by distributed, event-driven processing mechanisms that are massively parallel, resilient against failure or damage, and which consume comparably low energy (1-10 fJ/synapse)~\cite{Chicca_etal14,Thakur_etal18,b4}.
Neuromorphic computing mainly relies on collections of processing units, called neurons, consisting of synapses and somas.
Their main function is to integrate the synapse-weighted input signals and to produce an all-or-none event (a somatic spike) as soon as this integral exceeds a spiking threshold, which is then propagated to other neurons~\cite{Indiveri_etal11,b6,b35,b10}.

The synaptic circuits in these architectures play a vital role in the learning and memory formation mechanism \cite{b7}. Their main function is to convert the pre-synaptic voltage spike to a post-synaptic current, and to weight, or scale, the input signal. Furthermore, these synaptic circuits are considered crucial elements for future intelligent Brain-Machine Interfaces (BMI) to bridge the gap between biological and artificial neural systems \cite{Sharifhazileh_etal21,b8}.
Silicon-based technologies are currently the dominant realization methods to implement brain-like computing systems \cite{Furber_Bogdan20,Davies_etal18,Moradi_etal18}.
The silicon technology offers ultra-fast operational speeds ($\geq$\,GHz) and high-density devices, with mature fabrication processes that are precise and well understood \cite{b9}.
However, silicon-based implementations are expensive and complex, and crucially suffer from lack of biocompatibility, flexibility, and large area coverage.
Organic electronics and materials are an alternative to conventional electronics that can be integrated with low-temperature processes with relatively low-priced equipment over a large area.
Further advantages include ambipolar semiconducting behavior, physical flexibility, stretchability, and biocompatibility \cite{b11,b12,b34}.

Early proposals to emulate synaptic functions relied on multielement electric circuits \cite{Bartolozzi_Indiveri07a}. Since the announcement of a fabrication of a "memristor" \cite{b51}, there has been a great interest to employ these two-terminal inorganic or organic devices to emulate the function and the efficiency of biological synapses in a compact and simple form \cite{b5,b14,b52}.
However, they present a limited number of tunable parameters, typically ON/OFF resistance or discharge rates~\cite{b15}.
Multi-element synaptic circuits provide more flexibility at the cost of lower density~\cite{Chicca_Indiveri20}.
Despite the complexity of these circuits, compared to a single memristive device, multi-element synapse circuits offer control over individual parameters, provide continuously tunable weight, and enable the emulation of biophysically realistic synaptic temporal dynamics~\cite{b16,b17,b38,b39}.

One of the main characteristics of an ideal neuromorphic mechanism is having a biologically plausible time constant (in excess of tens of milliseconds) to process real-world sensory signals efficiently and interact with the environment in real time \cite{b18,b19,b20}.
 Log-domain subthreshold circuits with large capacitors faithfully provide biologically plausible temporal dynamics \cite{b21,b22,b24}.
Several log-domain synaptic circuits have been proposed \cite{Bartolozzi_Indiveri07a}.
In particular, the Log-Domain Integrator (LDI) synapse introduced by Merolla and Boahen in 2004 is a linear filter that lets the synapse integrate the contribution of action potentials from multiple sources linearly \cite{b25}.
The main drawback of the silicon-based circuit is that long time constants require significant silicon area, for the capacitor, which in turn reduces the number of synapses that can be integrated on a single die~\cite{b26}.

Organic materials are characterized by intrinsically slower charge carrier mechanisms compared with inorganic materials; therefore, the switching speeds of organic devices is limited below MHz or even kHz \cite{b27,b28}.
Integrating organic materials and LDI synapse architecture seems to offer an ideal synaptic circuit with a plausible time constant and a linear behavior.

In here, we demonstrate a physically flexible spiking LDI synapse fabricated using organic electronics on a flexible plastic substrate.
The circuit is fabricated using complimentary p- and n-type organic materials. Following the fabrication, the organic field-effect transistors (OFETs) and the LDI synapses are characterized and compared before and during bending. We demonstrate that the synaptic circuit converts pre-synaptic voltage spikes to post-synaptic current. We also show that the magnitude of the post-synaptic current is proportional to the synaptic strength, adjusted via the weighing voltage ($V_W$). The time constant of LDI synaptic circuits is experimentally estimated and compared while flat and under strain. The strain shifted the threshold voltage of p- and n-type OFETs by 0.63\,V and 1.01\,V, respectively. The estimated time constant with a 10\,nF synaptic capacitor reached 126\,ms and 221\,ms before and during bending.
Finally, the effect of disparity in capacitance, pre-synaptic signal, and weighting voltage are studied on the time constant under neutral and strain conditions.

\section{Materials and methods} 

\subsection{Device structure}

\begin{figure}[H]
\centering
\subfloat[]{\label{fig_1a}\includegraphics[height=2in]{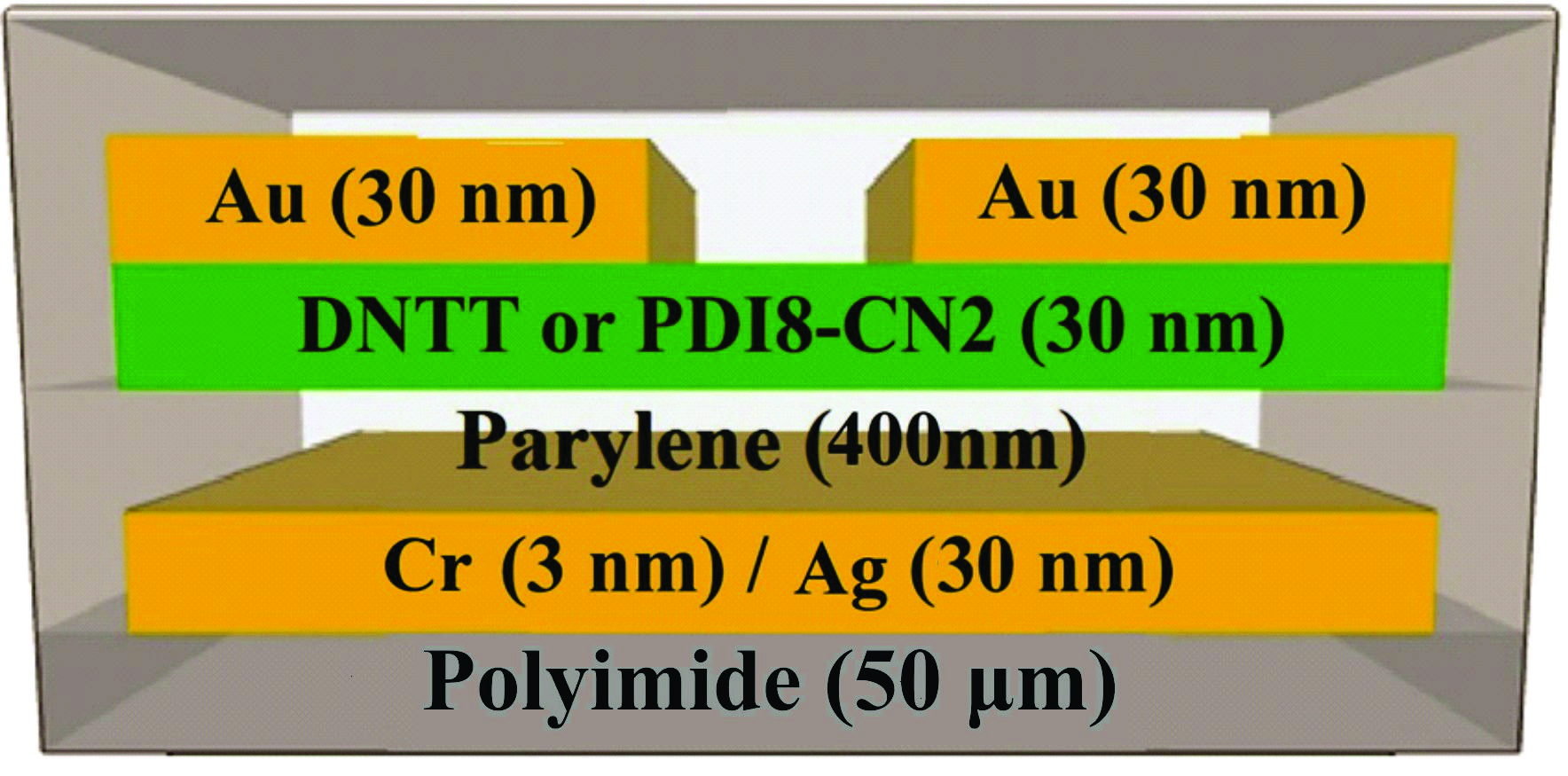}}
\hspace{10pt}
\subfloat[]{\label{fig_1b}\includegraphics[height=3in]{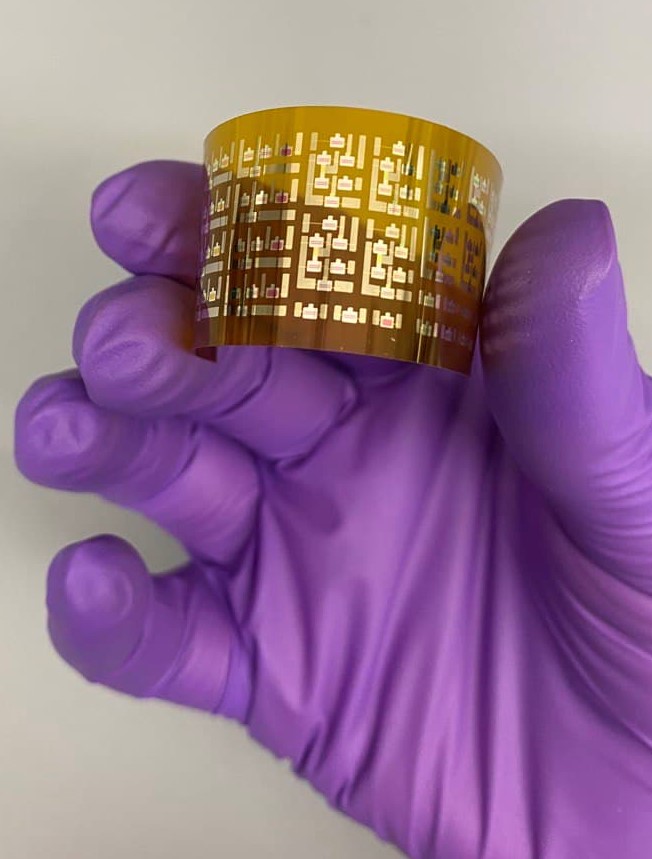}}
\hspace{10pt}
\subfloat[]{\label{fig_1c}\includegraphics[height=1in]{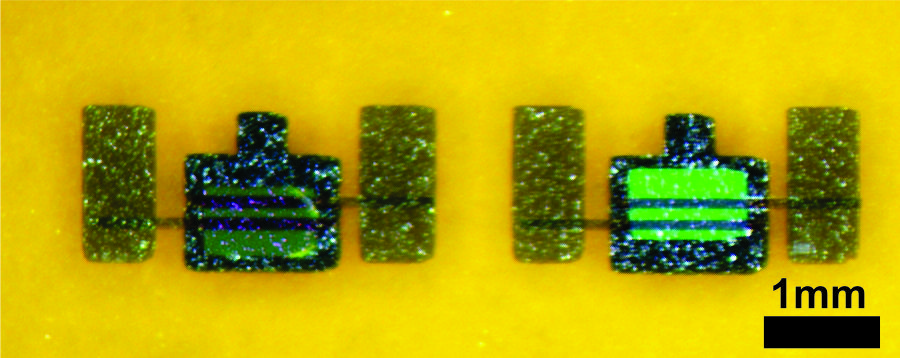}}
\caption{(a) Device structure of organic transistors.
(b) Photograph of the flexible chip with organic transistors and synaptic circuits.
(c) Zoomed view of n-type (left) and p-type (right) organic transistors fabricated side by side on the same substrate.
The bright dots are due to the structure of Polyimide film and do not negatively affect the devices' performances.}
\label{myfig1}
\end{figure}

The main elements of the LDI synaptic circuit are p- and n-type OFETs.
\textbf{Figure~\ref{myfig1}} shows the device stack with photographs of the flexible chip and both types of OFETs. The OFET structure is top-contact bottom-gate, consisting of the following layers: Polyimide (PI) substrate, Cr/Ag gate, Parylene diX-SR as gate dielectric, active layers of Dinaphtho[2,3-b:2',3'-f]thieno[3,2-b]thiophene (DNTT), and N,N'-bis(n-octyl)-x:y,dicyanoperylene-3,4:9,10-bis(dicarboximide) (PDI-8CN2, also referred to as N1200) for p- and n-type OFETs, respectively, and Au as source and drain.
The channel length and width are the same for both types of OFETs, at 100\,$\mu$m and 1000\,$\mu$m, respectively.

\subsection{ Characterization of organic transistors }

\begin{figure}[H]
\centering
\subfloat[]{\label{fig_2a}\includegraphics[height=2in]{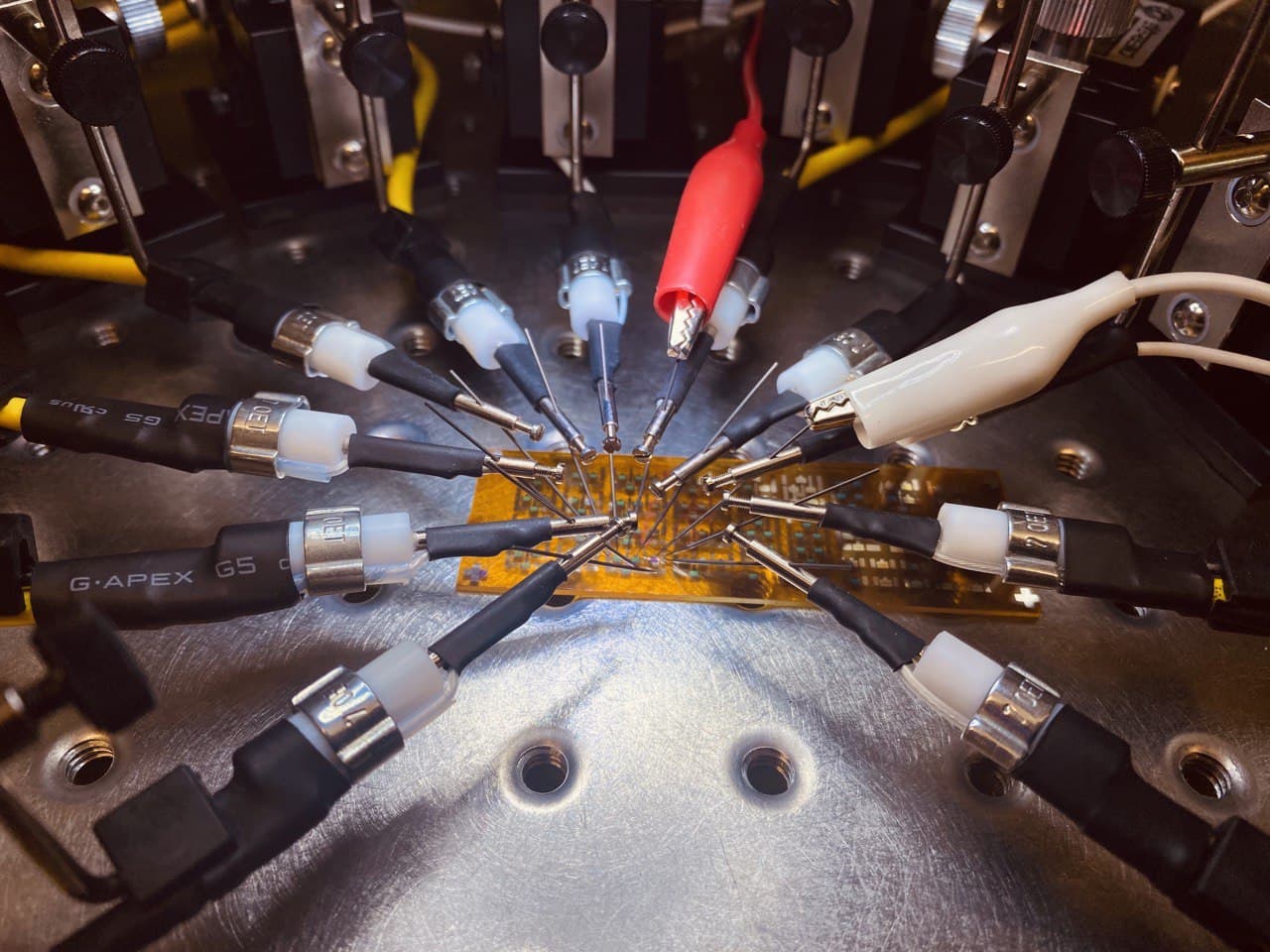}}
\hspace{10pt}
\subfloat[]{\label{fig_2b}\includegraphics[height=2in]{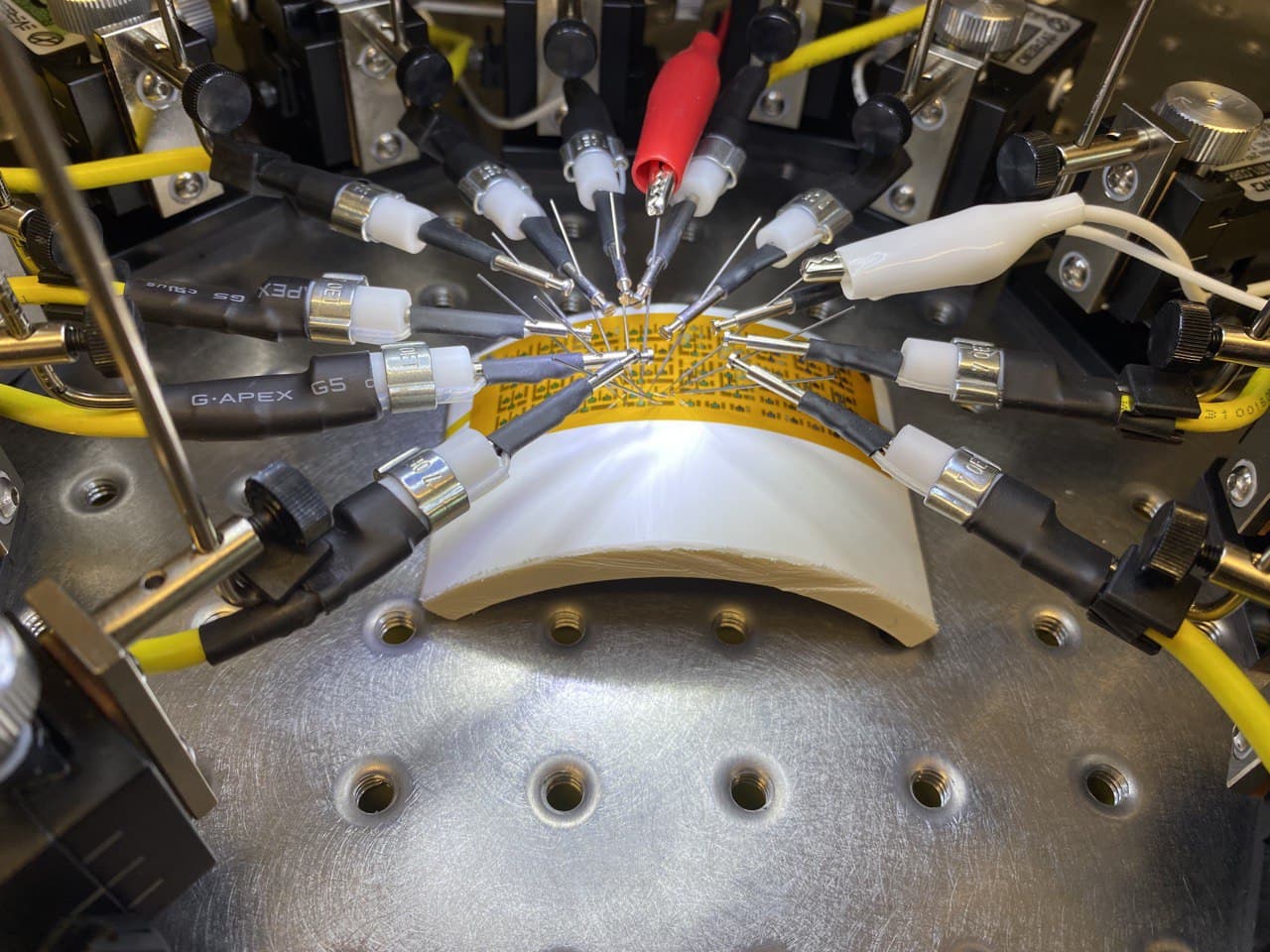}}
\caption{(a) The characterization setup before (flat) and (b) during bending, with individual micromanipulators used to access individual circuit nodes clearly visible.}
\label{myfig2}
\end{figure}

The output and transconductance characteristics of individual OFETs and the entire LDI synaptic circuits are examined before (flat) and during bending to elucidate the effects of stress and compare their electrical properties.
The bending radius is 4.5 inches (114.3\,mm), and \textbf{Figures~\ref{myfig2}} demonstrates the laboratory test setup, including individual micromanipulators used to directly contact individual circuit nodes (e.g.
$V_{DD}$, $I_{Syn}$, or $GND$).

\begin{figure}[H]
\centering
\subfloat[]{\label{fig_3a}\includegraphics[height=2in]{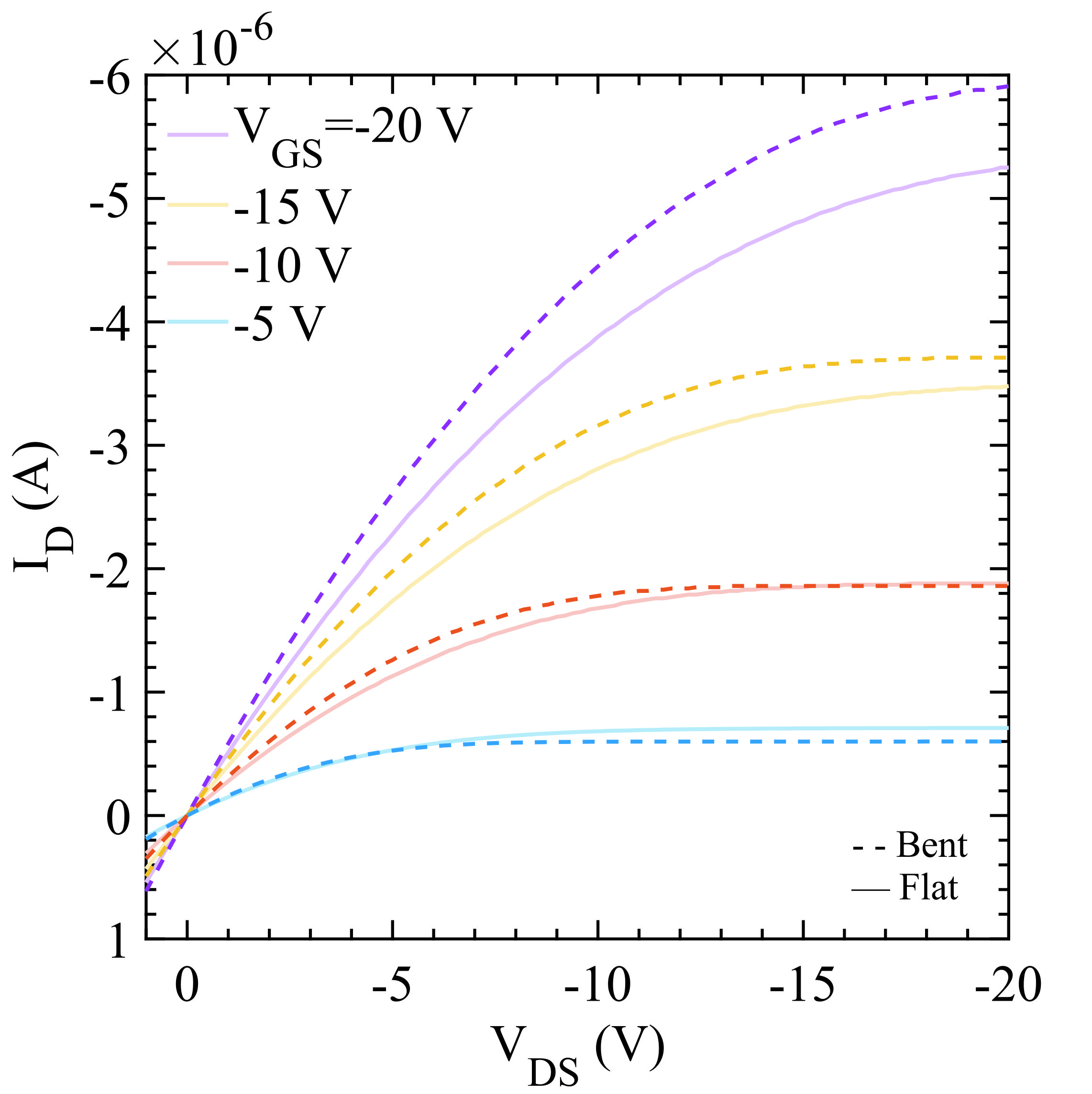}}
\hspace{10pt}
\subfloat[]{\label{fig_3b}\includegraphics[height=2in]{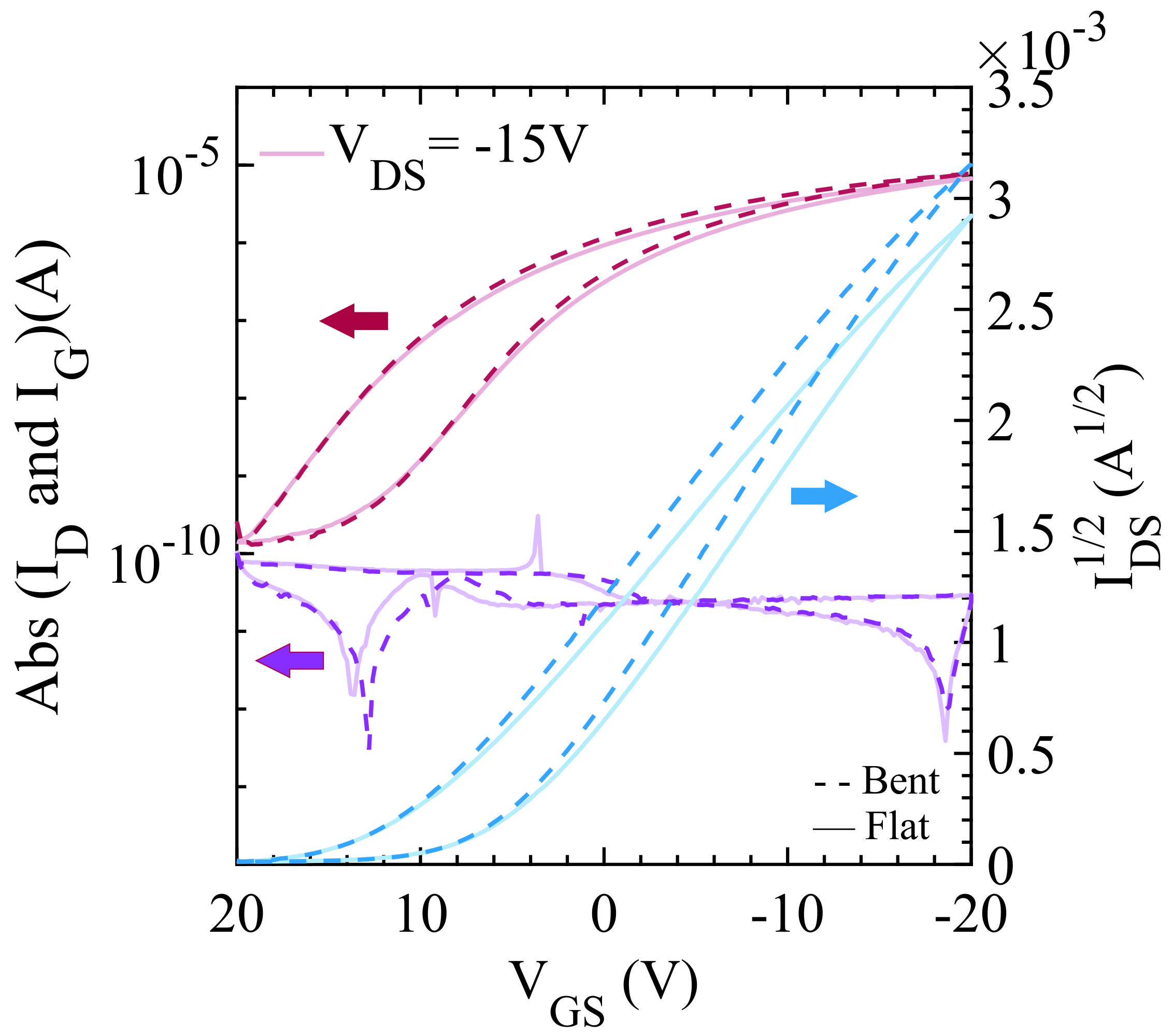}}
\caption{(a) Output and (b) transconductance curves of the \textbf{p-type} organic transistor before (flat, solid line) and during (bent, dashed line) bending to a radius of $4.5$ inches ($114.3$ \textit{mm}), shown in \textbf{Figure \ref{myfig2}}.}
\label{myfig3}
\end{figure}

\begin{figure}[H]
\centering
\subfloat[]{\label{fig_4a}\includegraphics[height=2in]{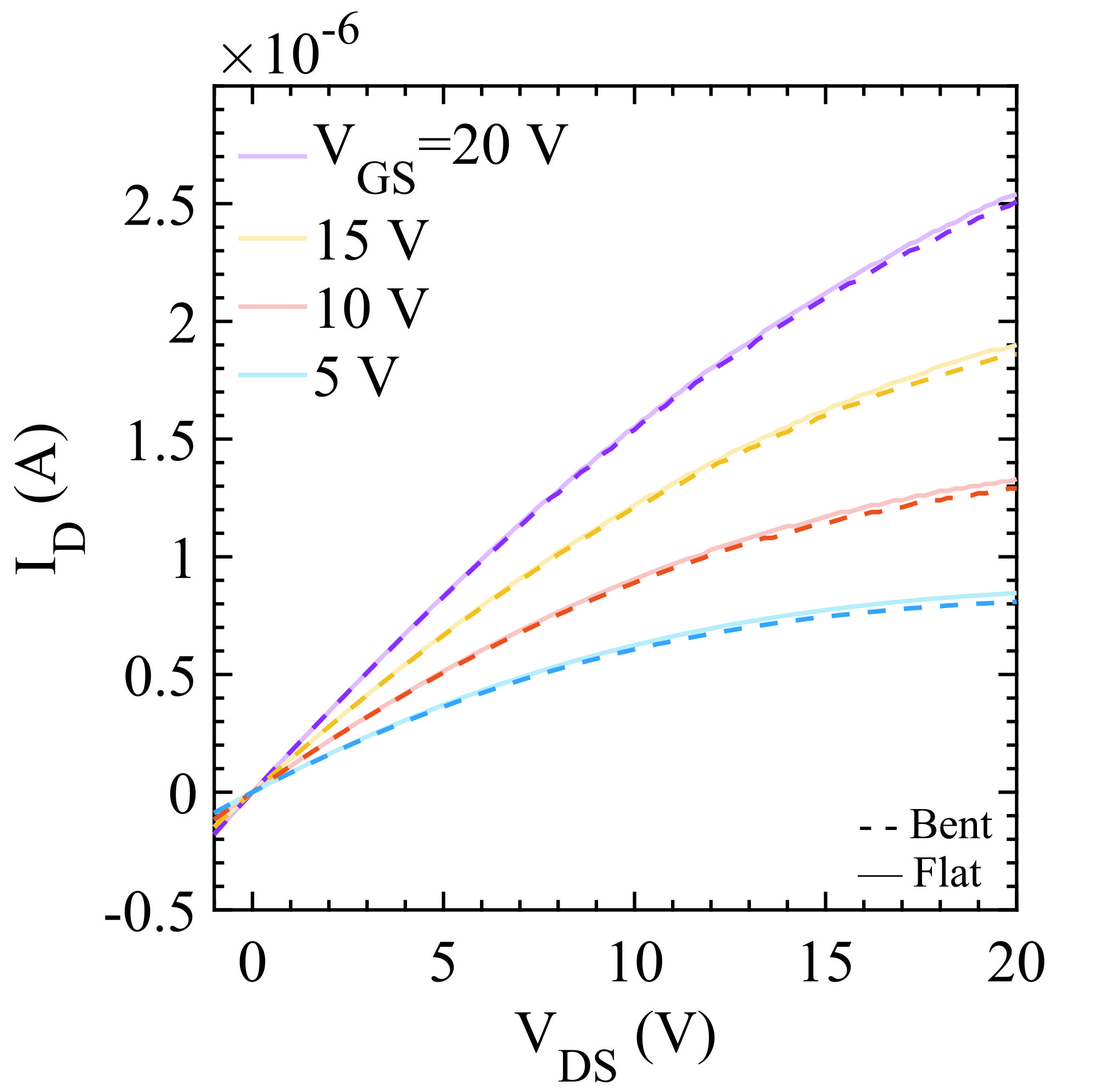}}
\hspace{10pt}
\subfloat[]{\label{fig_4b}\includegraphics[height=2in]{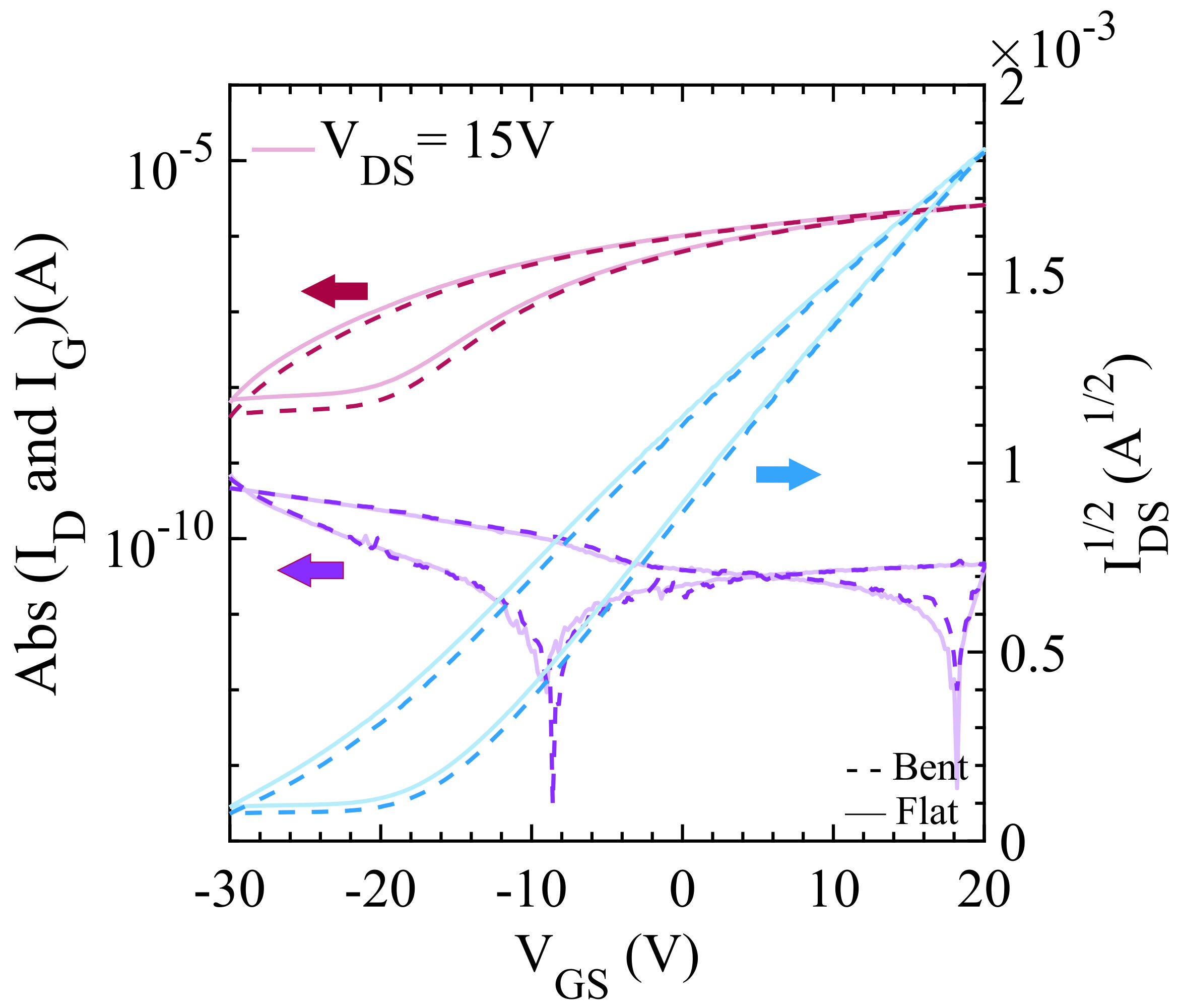}}
\caption{(a) Output and (b) transconductance curves of the \textbf{n-type} organic transistor before (flat, solid line) and during (bent, dashed line) bending to a radius of $4.5$ inches ($114.3$ \textit{mm}), shown in \textbf{Figure \ref{myfig2}}.}
\label{myfig4}
\end{figure}

\textbf{Figures~\ref{myfig3}} and \textbf{\ref{myfig4}} present representative examples of p- and n-type OFETs characterization results, shown with neutral and strain status with solid and dashed lines, respectively.
\textbf{Table~\ref{tab1}} presents the OFETs' characterization results in flat and bent conditions.

\begin{table}
\centering
 \caption{The characterization results of p- and n-type OFETs shown in \textbf{Figures~\ref{myfig3}} and \textbf{\ref{myfig4}}}
 \label{tab1}
  \begin{tabular}[h!]{@{}ccc@{}}
    \hline
      Parameters & Flat  & Bent \\
    \hline
    $V_{T}$ of p-type OFET (V)  & 5.35 & 5.98  \\
    \hline
    $V_{T}$ of n-type OFET (V)  & -19.39 & -18.4  \\
    \hline
    $\mu$ of p-type OFET (cm$^{2}$/Vs) & 0.31 & 0.34  \\
    \hline
    $\mu$ of n-type OFET (cm$^{2}$/Vs) & 0.050 & 0.052  \\
  \end{tabular}
\end{table}

Bending shifts the threshold voltage ($V_{T}$) of the p- and n-type OFETs by $0.63~V$ and $1.01~V$ towards more positive values and increases the mobility ($\mu$) by $0.03 ~cm^{2}/Vs$ and $0.002 ~cm^{2}/Vs$, respectively.
The carrier mobility of the n-type device is lower than the p-type, which decreases the switching speed of the device.
The OFF current of p-type device remains constant before and during bending, $ 1.54\times10^{-10}~A$ at $20~V$, while bending decreases the OFF current of n-type device by $2.4 ~nA$ at $-30~V$.

\subsubsection{ The mechanisms of hysteresis }
\textbf{Figures~\ref{fig_3b}} and \textbf{\ref{fig_4b}} show that $I_D$ depends on $V_{GS}$'s sweep direction, known as the "hysteresis" phenomenon. Such reversible electrical bistabilities are often observed in OFETs. A variety of effects have been identified as causes, including charge trapping at the interface of the semiconductor and the dielectric, the dielectric polarization, injection of charges from the semiconductor/gate to the dielectric bulk, moving ions in the dielectric, and slow reaction of moving charge carriers \cite{b47,b48}. While hysteresis has been used as the basis of memory devices \cite{b49}, generally it is seen as a negative
effect with adverse effects on the electrical circuits. As discussed later, it is likely one of the culprits of subtle but noticeable changes in our synaptic circuit, for instance affecting the time constant. Possible fabrication strategies aiming to minimize hysteresis include replacing the gate electrode, reducing the dielectric thickness, and adding Self-assembled Monolayers (SAMs) \cite{b50}.

\subsection{Log-domain integrator synapse}

The electrical neural signals are transmitted through synapses between individual neurons in the brain.
A human nervous system consists of approximately $10^{16}$ synapses that permit the signals to be transferred between neurons.
There are two types of biological synapses, electrical and chemical.
Chemical synapses tend to transmit more complicated signals than electrical ones.
Chemical synapses convert the electrical activities of a pre-synaptic neuron to the release of a chemical known as a neurotransmitter.
Neurotransmitters bind to receptors, mechanical elements in post-synaptic neurons, and initiate electrical activities that may either be inhibitory or excitatory.
Chemical synapses play a critical role in the formation of memory; therefore, there has been a considerable research effort focusing on emulating the synaptic functions \cite{b29}.

While several examples demonstrating functional performances have been introduced in the literature,  the log-domain integrator (LDI) synaptic circuit represents a biologically realistic current-mode model of a chemical synapse~\cite{Bartolozzi_Indiveri07a}.

\begin{figure}[H]
\centering
\subfloat[]{\label{fig_5a}\includegraphics[height=2in]{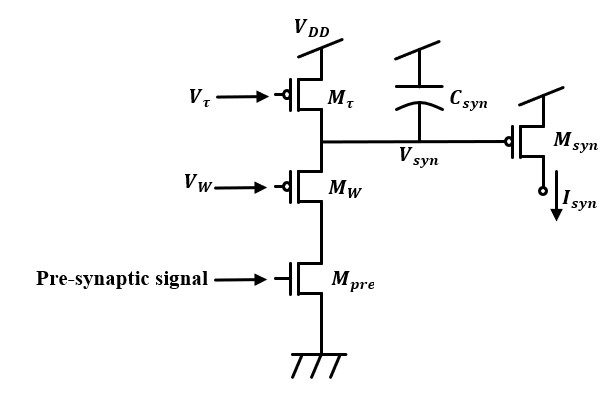}}
\hspace{10pt}
\subfloat[]{\label{fig_5b}\includegraphics[height=2in]{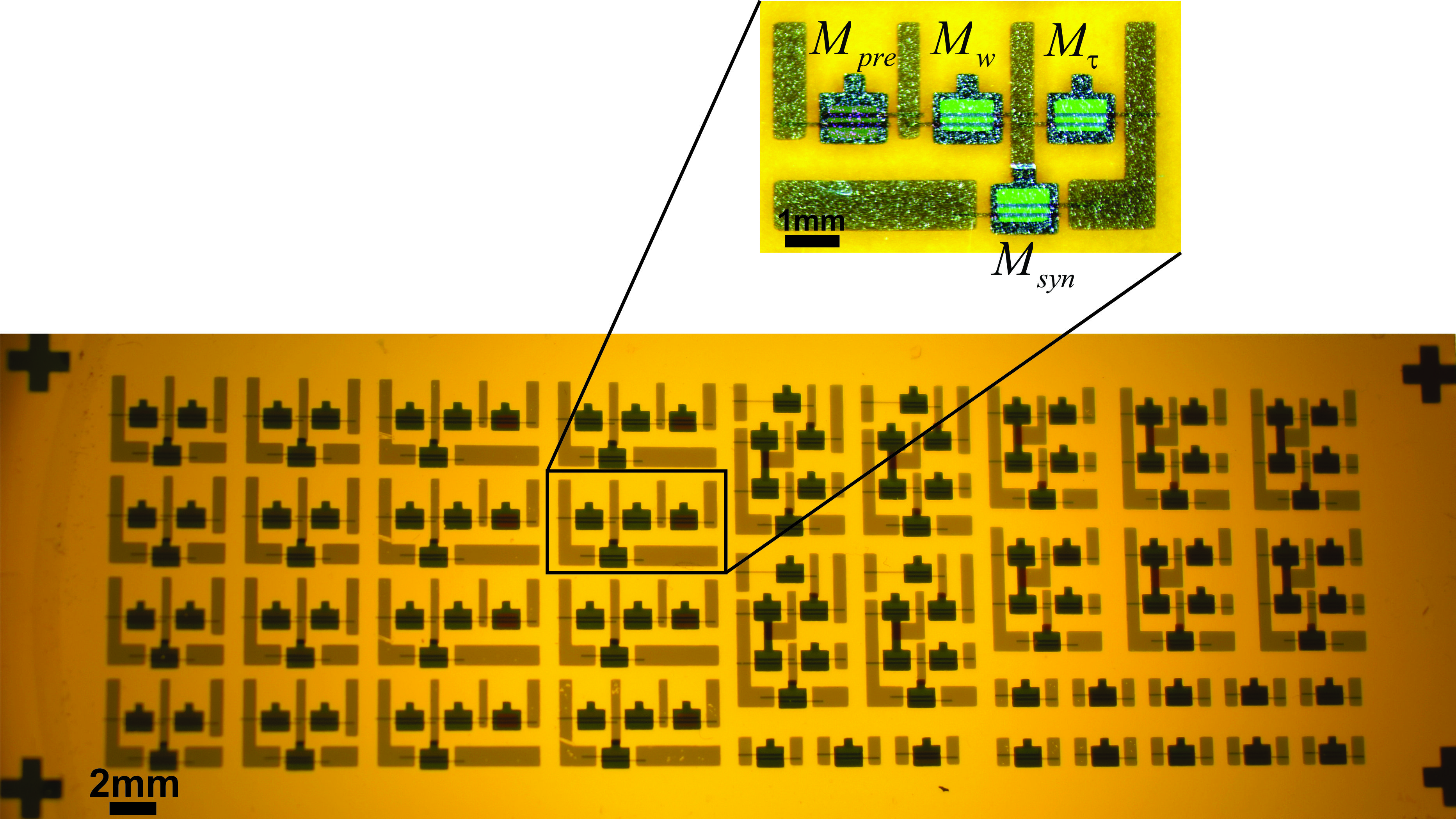}}
\caption{ (a) Circuit diagram of the implemented organic log-domain integrator synapse.
(b) Photograph of  an entire chip, with multiple organic circuits, with a zoomed view of a single log-domain integrator synapse on a Polyimide substrate shown above.}
\label{myfig5}
\end{figure}

\textbf{Figure~\ref{fig_5a}} illustrates the LDI circuit schematic. It consists of three p-type, one n-type OFET, and a capacitor.
In order for the circuit to function properly as a log-domain integrating circuits, all of the p-type OFETs need to operate in the subshreshold regime (otherwise the circuit works as a nonlinear and power-hungry reset-and-discharge synapse). 
However, the intrinsic characteristics of n-type OFETs (higher OFF current and lower mobility than p-type OFET) make subthreshold operation more challenging \cite{b42}.
The OFF current of the n-type $M_{pre}$ OFET (which acts as a switch turning the synaptic circuit ON or OFF) is greater than the OFF current of p-type OFETs by approximately one order of magnitude ($6.92 \times 10^{-9}~A$ compared with $1.54 \times 10^{-10}~A$).
  % Therefore, the n-type OFET ($M_{pre}$) remains ON compared to p-type when the OFETs operate in the subthreshold regime. 
Biasing p-type $M_{\tau}$ OFET in a $V_{GS}$ that produces an $I_D$ greater than the OFF current of n-type $M_{pre}$ OFETs allows for $M_{pre}$ to turn the circuit off, and the LDI synapse operates appropriately, as will be demonstrated in \textbf{Section~\ref{results}}. 
Assuming all the p-type OFETs  are working in a subthreshold regime, the LDI synaptic behavior can be described as follows:
a square pulse signal, which represents the pre-synaptic voltage spike, activates $M_{pre}$.
When the $M_{pre}$ is ON, the gate voltage of $M_{syn}$ decreases with a rate set by $I_w – I_\tau$, and the synaptic current ($I_{syn}$) increases following an exponential profile.
At the onset of each pre-synaptic pulse, the capacitor discharges, and the $I_w$ decreases exponentially.
When the input voltage pulse ends, the $M_{pre}$ is turned OFF, the capacitor $C_{syn}$ is linearly recharged to $V_{DD}$ by the constant current through $M_{\tau}$, and the current through $M_{syn}$ decreases back to its leakage current levels.
The parameter $V_W$ biases the gate of $M_{W}$ and sets the maximum efficacy of the synapse (i.e., the synaptic weight).
The $M_\tau$ transistor needs to be biased to produce a current that compensates for the leakage current of the n-type $M_{pre}$ to charge the synaptic capacitor $C_{syn}$.

\textbf{Figure~\ref{fig_5b}} shows a photograph of the fabricated organic synapse.
Individual OFETs are connected through $30 ~nm$ Au tracks immediately after source/drain deposition.
A non-circuit-integrated commercial capacitor is deployed to expedite the fabrication and characterization process, as an integrated organic capacitor (part of our future efforts) would dramatically increase the complexity of the fabrication.

\section{Results}

\subsection{Synaptic circuit characterization}\label{results}

\begin{table}
\centering
 \caption{Experimental parameters with data graphically shown in \textbf{Figure~\ref{myfig6}}}
 \label{tab2}
  \begin{tabular}[!H]{@{}ccccc@{}}
    \hline
     $V_W$ (V) & $C_{syn}$ (nF) & Pre-synaptic signal period/width (s) & $V_\tau$ (V)& $V_{DD}$ (V)\\
    \hline
    $9$, $10$, $11$  & $10$  & $4/2$  & $9$  & $15$\\
    \hline
  \end{tabular}
\end{table}

\begin{figure}[H]
\centering
\includegraphics[height=3in]{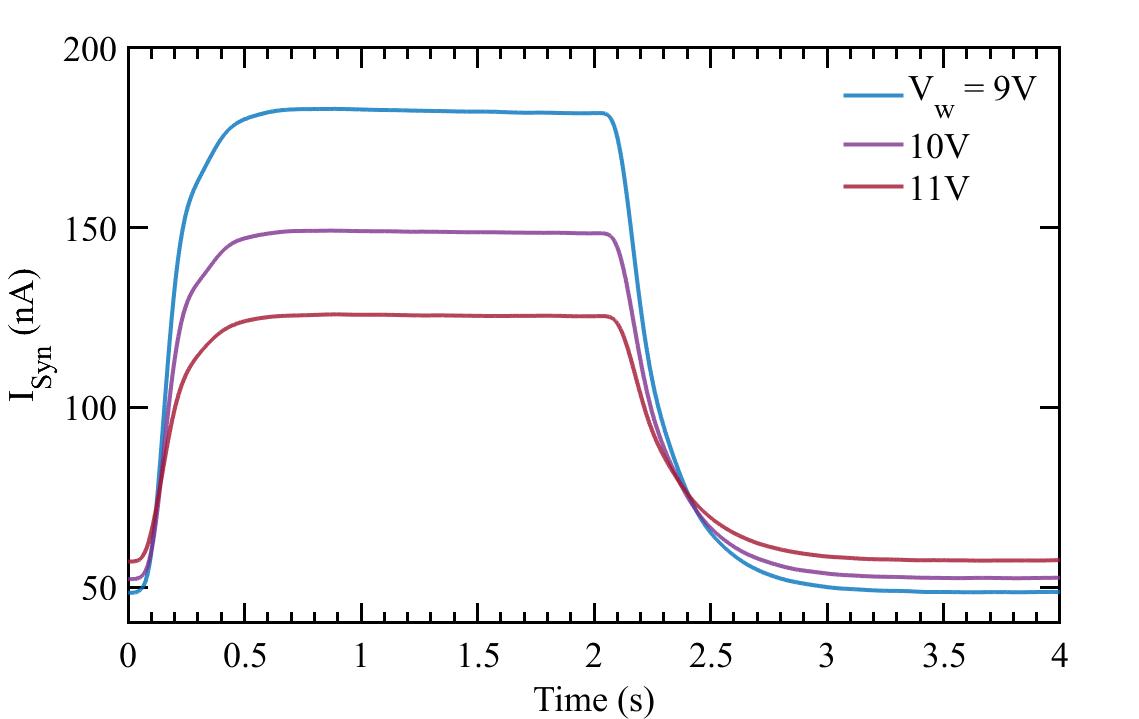}
\caption{ Step response of the log-domain integrator synapse for three different values of $V_W$.}
\label{myfig6}
\end{figure}

As mentioned, one of the functions of a synapse is to weight, or scale, the pre-synaptic input signal via synaptic weights. \textbf{Figure~\ref{myfig6}} illustrates the step response of the organic LDI synapse to a square wave with a cycle duration of 4 seconds, alternating between -10\,V and 10\,V, to turn OFF and ON $M_{pre}$ based on the n-type OFET characterization results.
The circuit's response has been plotted for three synaptic weights ($V_W$) to demonstrate the circuit's functionality in either signal attenuation or amplification.
$M_\tau$ needs to produce a current greater than the OFF current of n-type $M_{pre}$ OFET, while $V_{DS}$ of $M_\tau$ is a value less than 1\,V.
Therefore, 9\,V is applied to the gate of $M_\tau$ with $V_{DD}$=15\,V to compensate the leakage current of $M_{pre}$ when pre-synaptic voltage is -10 V.
\textbf{Table~\ref{tab2}} summarized the experimental values.

The synaptic weight parameter $V_W$ modulates the height of the circuit's response, namely the magnitude of the saturated synaptic current $I_{syn}$.
Specifically, applying smaller $V_W$ values decreases $V_{syn}$, consequently elevating the peak of the output synaptic current ($M_{W}$ is a p-type OFET and the source of $M_{W}$ is connected to $V_{syn}$. Applying lower $V_{W}$ means smaller $V_{GS}$ ($V_{W}$ - $V_{syn}$). If we consider that $V_{syn}$ is almost constant with a value close to $V_{DD}$, then $V_{GS}$ of $M_{W}$ depends only of $V_{W}$. Hence smaller $V_{W}$ means higher $I_{W}$. Consequently, $C_{syn}$ can recharge more and $V_{GS}$ of $M_{syn}$ varies more and produce taller peaks of $I_{syn}$).
During the step input, while $M_{pre}$ is ON, $M_\tau$ produces a current that is approximately constant.
As soon as the pre-synaptic voltage is turned to -10\,V, $I_\tau$ starts charging $C_{syn}$, eventually turning off both $M_{syn}$ and $M_\tau$.
The threshold voltage of p-type OFETs is a non-zero value in the characterizations; therefore, the $M_{syn}$ is not turned off, and $I_{syn}$ is biased to a constant value greater than zero.
Either applying a voltage less than $V_{DD}$ to the source electrode of $M_{syn}$ or fabricating a p-type OFET with a threshold voltage close to $0 ~V$ will remove the bias from the result.\\
Ideally, the synaptic currents need to reach the same steady-state value with different $V_W$s; however, \textbf{Figure~\ref{myfig6}} shows the $I_{syn}$s are marginally different at the end of the cycle (t $<$ 3 s). Weighting voltages stimulate hysteresis mechanisms and lead to a discrepancy in the steady-state values.

\subsection{Time constant} 
Leaky integrate-and-fire neurons can distinguish between different temporal input spike patterns only if the synapses stimulated by the input spike patterns exhibit dynamics with time constants comparable to the time constant of the neuron's membrane potential \cite{b26}. As such, synaptic circuits with time constants of milliseconds or seconds are critical.\ Silicon synaptic circuits have shown the same time constants as this study, but with smaller capacitors in the range of pico to femtofarads—partially due to the superior inorganic semiconducting technologies, such as carrier mobilities, lower OFF currents, and matching threshold voltages of p- and n-type devices.
Moreover, organic semiconducting technology is relatively new compared with mature inorganic electronics.
Therefore, further improvement in the fabrication of this study, regarding materials, deposition, and patterning methods, will yield organic synaptic circuits with a large time constant using smaller capacitors.
Finally, organic materials naturally offer biocompatibility and flexibility, which are difficult or impossible for silicon technologies.

%\threesubsection{First lowest-level subsection}

\subsubsection{Experimental time constant}\label{Exp}

The governing equations of the LDI synaptic circuit are only valid when all p-type OFETs are operating in subthreshold regimes.
However, because our p-type OFETs operate in a weak-inversion regime, resulting in a quasi-linear circuit operation, the standard equations cannot be used directly to extract the circuit's time constant.

LDI synaptic circuit implements a first-order low-pass filter; therefore, the time constant can be extracted from circuit's step response \cite{b16}.
Regardless of the transistors' operating regime, \textbf{Figure~\ref{myfig6}} shows that the organic LDI synapse still works similarly to a first-order low-pass filter (the non-ideal behavior of the circuit seen as a difference in OFF currents, is likely caused by the hysteresis effect, which also contributes to a slight shift to the current overtime).
The time constant can be experimentally estimated through the circuit's step response.
Various methods exist to estimate the time constant of a first-order system \cite{b43,b44,b45,b46}.
Indiveri et al. fitted the experimental data with an exponential equation to estimate the time constant \cite{b26}. \textbf{Equation~\ref{eq_1}} presents the exponential relationship between the synaptic current and the time constant. The fitting parameters are estimated using a Particle Swarm Optimization algorithm.\
\begin{equation} \label{eq_1}
I_{syn} =
    \begin{cases}
      a+b \times e^{\frac{-t}{\tau}} & \text{charge phase}\\
      c \times e^{\frac{-t}{\tau}} & \text{discharge phase}\\
    \end{cases}       
\end{equation}

\subsubsection{The effects of synaptic capacitance on the time constant}\label{sec1}

\begin{table}
\centering
 \caption{Experimental parameters discussed in \textbf{Section~\ref{sec1}}}
 \label{tab3}
  \begin{tabular}[H!]{@{}ccccc@{}}
    \hline
    $V_W$ (V) & $C_{syn}$ (nF) & Pre-synaptic signal period/width (s) & $V_\tau$ (V)& $V_{DD}$ (V)\\
    \hline
   $10$  & $4.7$,$~10$  &$2/1$, $1/0.5$  & $9$  & $15$\\
    \hline
  \end{tabular}
\end{table}
\vspace*{1 cm}
\begin{table}
\centering
 \caption{Statistical information graphically shown in \textbf{Figure~\ref{myfig8}} for $C_{Syn}=4.7~nF$ and $V_W=10~V$}
 \label{tab4}
 \centering
  \begin{tabular}[!htbp]{@{}cccccc@{}}
    \hline
       Condition &$Period_{presyn}~(s)$& \emph {Min} (s) & \emph {Max} (s) & \emph {Median} (s) &\emph {Mean} (s) \\
    \hline
    		Flat&	2	&67.15	&70.58	&68.21	&68.50\\
    \hline
    	Bent	&2	&91.67	&105.19	&97.08	&97.28\\
    \hline
   	    Flat	&1	&66.45	&68.21	&67.03	&67.08\\
    \hline
        Bent	&1	&79.55	&97.28	&89.25	&89.15\\

  \end{tabular}
\vspace*{1 cm}
\centering
 \caption{Statistical information graphically shown in \textbf{Figure~\ref{myfig9}} for $C_{Syn}=10~nF$ and $V_W=10~V$}
 \label{tab5}
 \centering
  \begin{tabular}[!htbp]{@{}cccccc@{}}
    \hline
       Condition &$Period_{presyn}~(s)$& \emph {Min} (s) & \emph {Max} (s) & \emph {Median} (s) &\emph {Mean} (s) \\
     \hline
    		Flat&	2&	121.01	&125.18	&122.97	&122.91\\
    \hline
    		Bent&	2&	157.45	&221.84	&191.14	&189.33\\
    \hline
    		Flat&	1&	107.15	&109.71	&108.40	&108.42\\
    \hline
    		Bent&	1&	105.17	&169.02	&144.90	&140.85\\
  \end{tabular}
\end{table}
The time constant is non-linearly proportional to the synaptic capacitance when the p-type OFETs are not operating in the subthreshold regime \cite{b31}.
Two capacitors, $4.7$ and $10 ~nF$, are deployed to show the effects of synaptic capacitance on the time constant.
In order for let $M_\tau$ produces a current greater than the leakage current of $M_{pre}$ when pre-synaptic voltage is $-10~V$, $9~V$ is applied to the gate electrode with $V_{DD} = 15~V$.
A square wave alternated between $\pm10~V$ with two different time periods—1 and 2 seconds—to simulate pre-synaptic voltage spikes.
\textbf{Table~\ref{tab3}} shows a summary of the experimental parameters.
The time constant is estimated for a captured cycle; therefore, pre-synaptic stimulations determine the number of estimated time constants in an experiment. In this experiment, 18 cycles are captured to estimate time constants. Whisker plots display patterns of estimated time constants in experiments.
 
\begin{figure}[H]
\centering
\label{fig_8a}\includegraphics[height=2.5in]{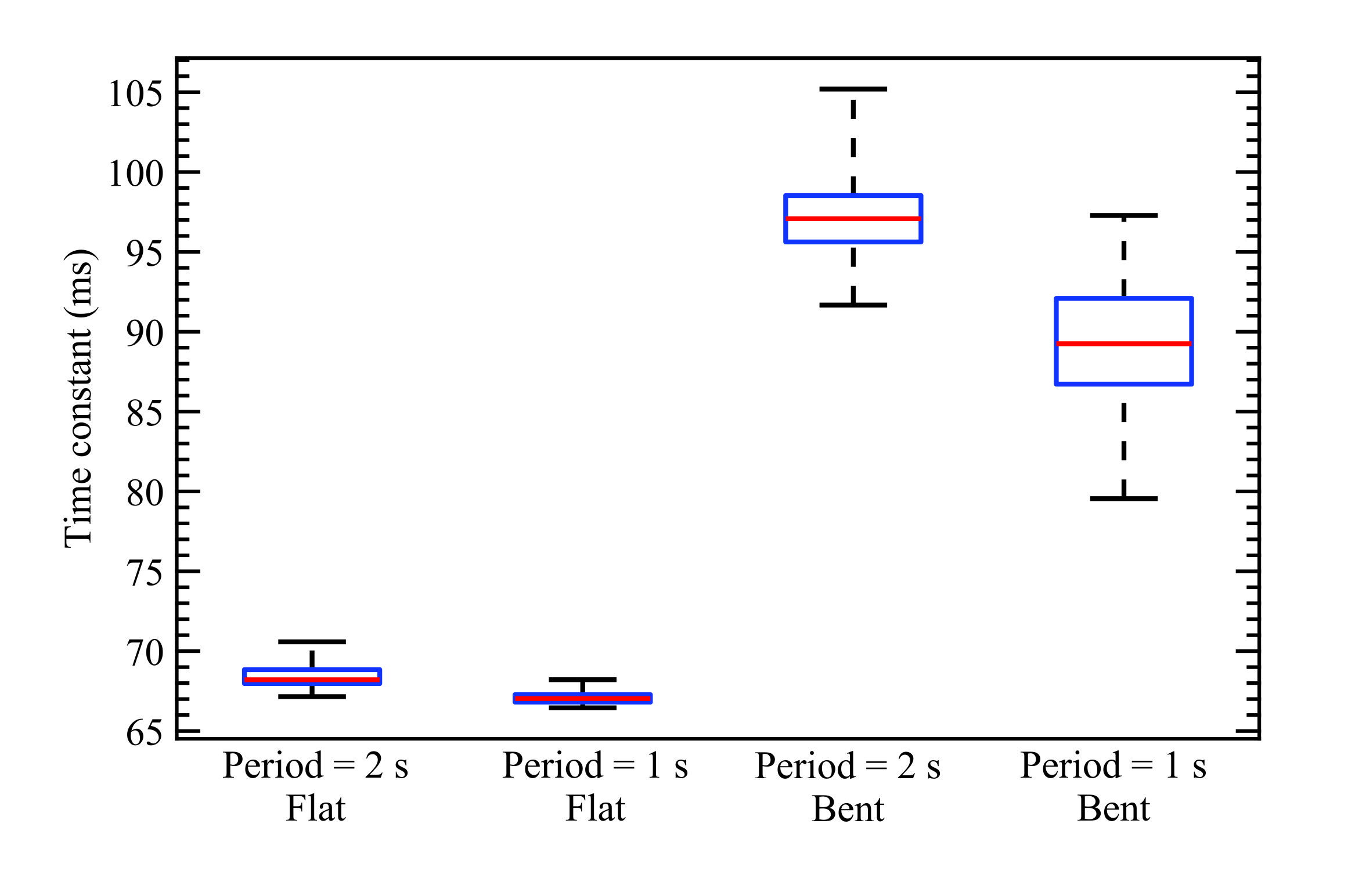}
\caption{Box plots of the estimated time constants for the log-domain integrator synapse with $C_{syn}=4.7~nF$ discussed in \textbf{Section~\ref{Exp}} to study the effect of synaptic capacitance and pre-synaptic signal periods.} 
\label{myfig8}
\end{figure}

\begin{figure}[H]
\centering
\label{fig_9a}\includegraphics[height=2.5in]{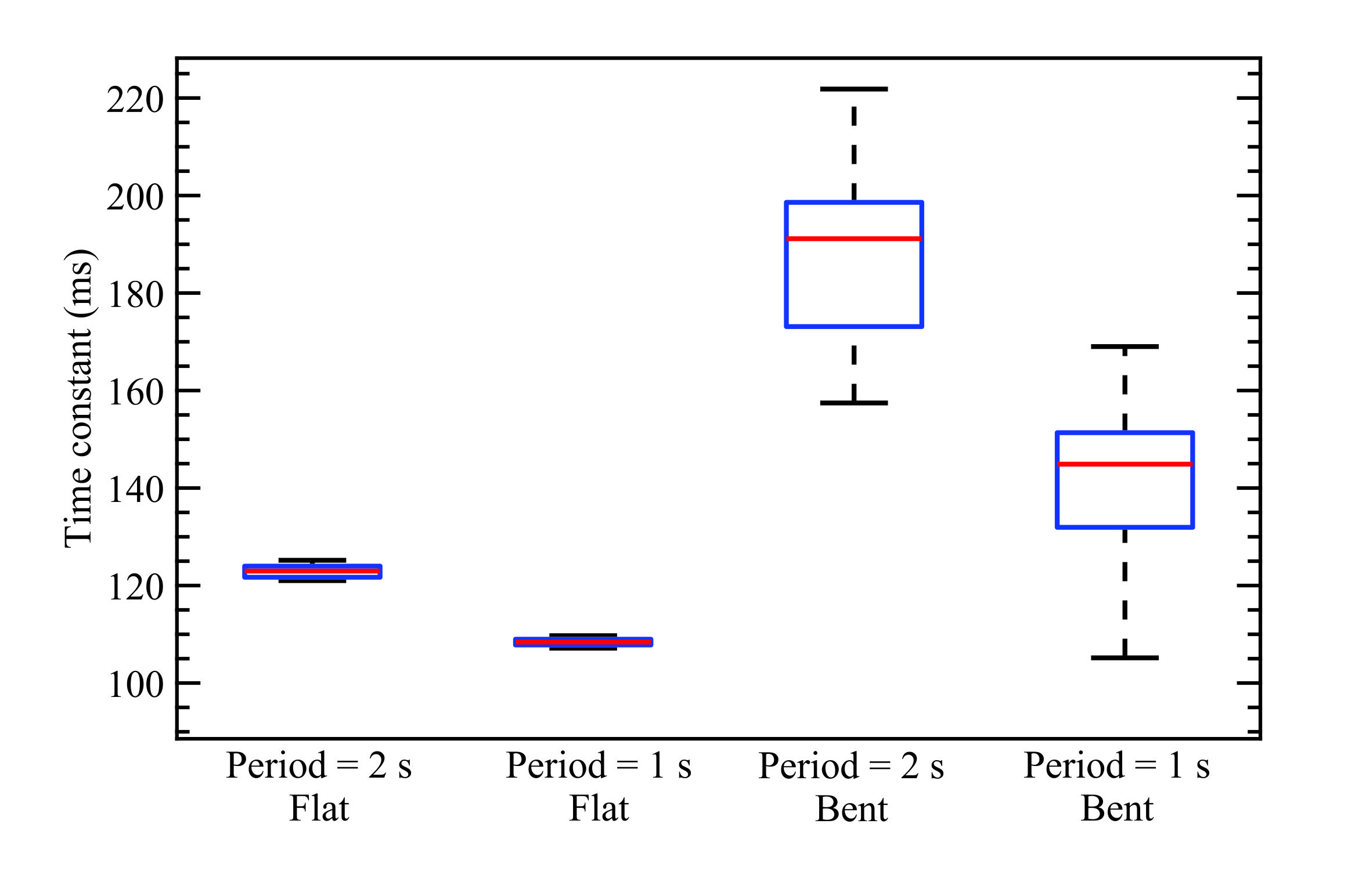}

\caption{Box plots of the estimated time constants for the log-domain integrator synapse with $C_{syn}=10~nF$ discussed in \textbf{Section~\ref{Exp}} to study the effect of synaptic capacitance and pre-synaptic signal periods}
\label{myfig9}
\end{figure}

\textbf{Figures~\ref{myfig8}} and \textbf{\ref{myfig9}} show the Box plots of estimated time constants based on \textbf{Section~\ref{Exp}} for $4.7$ and $10 ~nF$ synaptic capacitances.
The time constant has been estimated for two periods of pre-synaptic signals for every synaptic capacitance.
The time constant is independent of pre-synaptic signal periods and remained relatively unchanged for different periods of pre-synaptic signals before bending; however, during bending, the time constant increased due to the shifts in the threshold voltages.
Also,  the figures show disparities in medians and average values of time constants for the same conditions at the same synaptic capacitance for different pre-synaptic signal periods.
This is likely due to three factors.
Firstly, the hysteresis mechanisms affect the value of the time constant during the captured periods.
Secondly, the time constant estimation is intrinsically an error-prone process.
Finally, the system is not precisely a first-order low-pass filter but a higher-order system with the first-order dynamics as the leading dynamic.

\textbf{Table~\ref{tab4}} and \textbf{Table~\ref{tab5}} summarize the statistical information presented in \textbf{Figures~\ref{myfig8}} and \textbf{\ref{myfig9}}, respectively.

Supplemental \textbf{Figures~\ref{myfig12}} to \textbf{\ref{myfig15}}, shown in Supporting Information section, demonstrate the synaptic current ($I_{syn}$) for the experiments shown in \textbf{Section~\ref{sec1}}.

\begin{table}
\centering
 \caption{Experimental parameters shown in \textbf{Section~\ref{sec2}}}
 \label{tab6}
  \begin{tabular}[!htbp]{@{}ccccc@{}}
    \hline
    $V_W$ (V) & $C_{syn}$ (nF) & Pre-synaptic signal period/width (s) & $V_\tau$ (V)& $V_{DD}$ (V)\\
    \hline
    $9.5$, $9.8$ & $10$  & $4/2$, $2/1$ & $9$  & $15$\\
    \hline
  \end{tabular}
\vspace*{1 cm}

\centering
 \caption{Statistical information regarding \textbf{Figures~\ref{myfig10}} and \textbf{\ref{myfig11}} for $C_{Syn}=10~nF$}
 \label{tab7}
 \begin{tabular}[!htbp]{@{}ccccccc@{}}
    \hline
      $V_{W}$ (V)&Condition &$Period_{presyn}~(s)$& \emph {Min} (s) & \emph {Max} (s) & \emph {Median} (s) &\emph {Mean} (s) \\
     \hline
   	9.5&	Flat&	2&	122.46&	126.38&	124.23&	124.12\\
    \hline
    	9.5&	Bent&	2&	155.40&	179.64&	167.66&	167.43\\
    \hline
    	9.8&	Flat&	2&	115.98&	121.09&	119.07&	118.84\\
    \hline
    	9.8&	Bent&	2&	158.64&	198.24&	170.57&	174.02\\
    \hline
    	9.5&	Flat&	4&	119.99&	125.01&	122.16&	122.25\\
    \hline
    	9.5&	Bent&	4&	172.68&	185.13&	177.82&	178.73\\
    \hline
        9.8&	Flat&	4&	121.86&	125.29&	123.92&	123.71\\
    \hline
    	9.8	&Bent&	4&	160.99&	207.77&	180.75&	182.49\\
   \end{tabular}
\end{table}

\subsubsection{The effects of disparity of weighting voltage and period of pre-synaptic signal on the time constant}\label{sec2}

\begin{figure}[H]
\centering
{\includegraphics[height=2.5in]{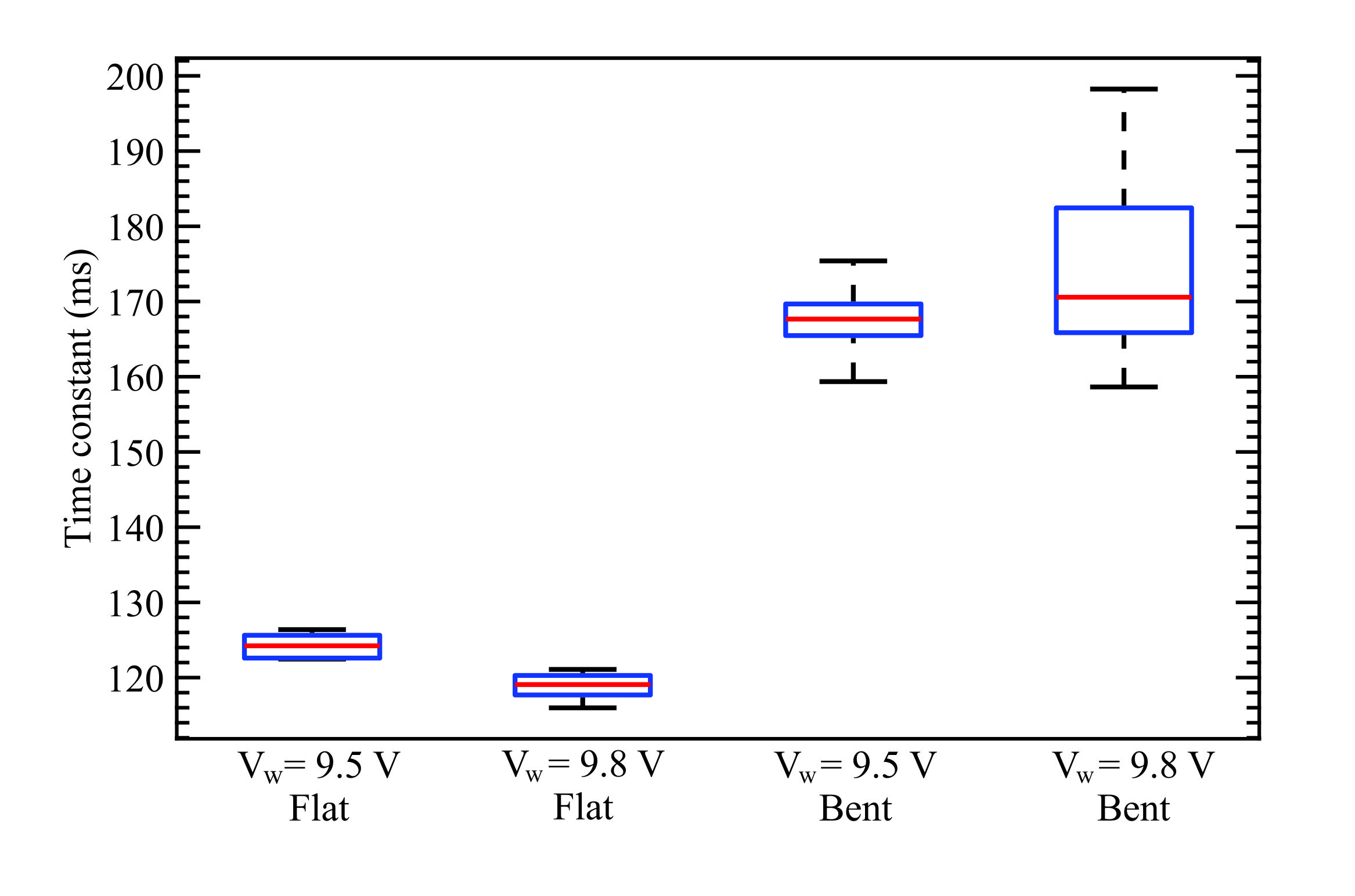}}
\caption{Box plots of the estimated time constant for the log-domain integrator synapse with $C_{syn}=10~nF$ and $Period_{psyn}=2~s$ discussed in \textbf{Section~\ref{Exp}} to study the effect of weighting voltage and pre-synaptic signal period.}
\label{myfig10}
\end{figure}

\begin{figure}[H]
\centering
{\includegraphics[height=2.5in]{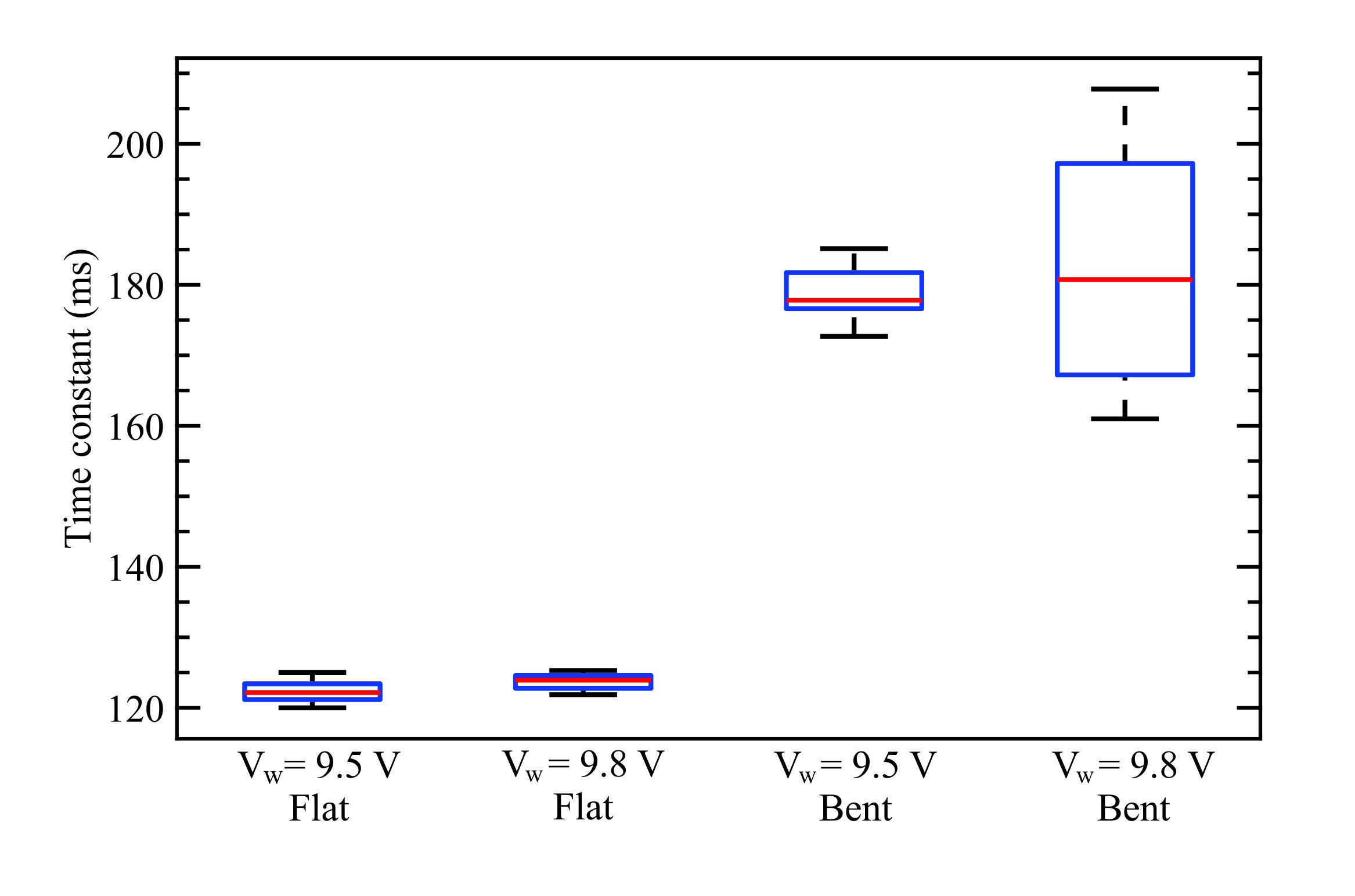}}
\caption{Box plots of the estimated time constant for the log-domain integrator synapse with $C_{syn}=10~nF$ and $Period_{psyn}=4~s$ discussed in \textbf{Section~\ref{Exp}} to study the effect of weighting voltage and pre-synaptic signal period.}
\label{myfig11}
\end{figure}

The weighting voltage and pre-synaptic signals have no role in determining the time constant; however, they do affect the saturation level of synaptic current.
Two different weighing voltages ($V_{W}$ of $9.5~V$, $9.8~V$) and square wave shape pre-synaptic signal with two different time periods—2 and 4 seconds—have been applied to the LDI synaptic circuit.
The experimental parameters are similar to the experiments discussed in \textbf{Section~\ref{sec1}} except the capacitance is constant ($10~nF$), and the weighing voltage and the periods of the pre-synaptic signal are varied.
\textbf{Table~\ref{tab6}} summarizes the experimental parameters. In this experiment, 18 cycles are captured to estimate time constants.
Whisker plots are deployed to show the statistical information regarding the experiment.
\textbf{Figures~\ref{myfig16}} and \textbf{\ref{myfig17}} show the synaptic currents with respect to time according to the parameters in the \textbf{Table~\ref{tab6}} before and during bending.
\textbf{Figures~\ref{myfig10}} and \textbf{\ref{myfig11}} present the box plots of experimentally estimated time constants according to \textbf{Section~\ref{Exp}} for $10 ~nF$ synaptic capacitance with pre-synaptic periods of two and four seconds and two weighing voltages.
It can be seen that the time constant is independent of weighing voltage and period of the pre-synaptic signal, with noticeable changes between flat and bent devices. However, the results show disparities in the results for different weighing voltages.
Also, the time constants need to be the same as what is shown in \textbf{Figure~\ref{myfig9}} to show the independency of time constant from periods of pre-synaptic signals.
The negligible disparity is mainly due to three reasons: 1) the hysteresis mechanisms that affect the time constants in a period of time, 2) the time constant extraction method that estimates the parameter is intrinsically susceptible to error, and 3) the higher-order dynamics that affect the response.
The threshold voltage shift caused an increase in the mean value of the time constant during bending. The statistical information regarding \textbf{Figures~\ref{myfig10}} and \textbf{\ref{myfig11}} have been summarized in \textbf{Table~\ref{tab7}}.

\section{Conclusion}

Synapses have two critical roles in neuromorphic systems, as individual interfaces with biological elements and forming abilities such as learning, memory, and cognition. Organic synaptic circuits offer advantages over silicon-based ones, including biocompatibility, flexibility, and large area covering. More importantly, the charge carrier mobilities of organic materials are slower than inorganic semiconductors, resulting in more plausible time constants. Furthermore, compared with individual devices used to emulate artificial synapses (e.g. two-terminal memristive devices), the multi-element synaptic circuits provide for a greater synaptic control, for instance via a continuously tunable weight or time constants.\hfill \break

There are two functions of a synapse. The first one is to convert pre-synaptic voltage spikes onto post-synaptic current. The second one is to scale, up or down, the magnitude of post-synaptic current according to so-called "synaptic weight" (which is adjusted during learning or training). We have demonstrated that our organic spiking synapse performs both of these functions, including the effects of synaptic weighting voltage $V_W$ and its effects on the post-synaptic current $I_{syn}$. This paper presents a biologically realistic current-mode model, linear charge and discharge, flexible organic log-domain integrator synapse consisting of three p-type, one n-type OFETs, and a capacitor. We also show that the time constant, estimated via fitting circuits step response, can reach $126~ms$ and $221~ms$ before and during bending when a $10~nF$ capacitor is deployed. We acknowledge large time constants can only be achieved using large capacitors. However, future improvements in individual device performance should lead to more practical, all-integrated solutions. The upgrades consist of 1) shrinking the dielectric thickness in the level of sub-100 nm to reduce the operating voltage, 2) lowering the OFF current to the level of picoamps, and 3) pushing threshold voltages toward zero for both p-type and n-type OFETs to remove any biases in the synaptic current. While these current results are not ready to be interfaced with biological systems, the research outcome opens the door to more biologically plausible time constant and biocompatible synaptic circuits, as well as networks of fully organic spiking neurons.
% Experimental section

\section{Experimental section}

\threesubsection{Materials and methods}\\

Dinaphtho[2,3-b:2',3'-f]thieno[3,2-b]thiophene (DNTT), as the p-type organic semiconductor, and N,N'-bis(n-octyl)-x:y,dicyanoperylene-3,4:9,10-bis(dicarboximide) (PDI8-CN2, also referred to as N1200), as the n-type semiconductor are obtained from Sigma-Aldrich and Polyerra, respectively.
Parylene diX-SR is purchased from Daisan Kasei and grown using a chemical vapor deposition (CVD) process with SCS LabCoater 3 (PDS 2010).
Chromium (Cr) rods, gold (Au), and silver (Ag) pellets are obtained from Kurt J.
Lesker Company (KJLC).
A NANO 36 thermal evaporation thin film deposition system by KJLC is exploited to deposit gate, source, drain, and active layers.
An HP 4155A performed the I-V measurements.
The flexible Polyimide substrates—$50 \mu m$ thick, pre-cleaned, $75\times 50 mm$—are obtained from DuPont de Nemours, Incorporation. A National Instrument USB-6343 Data Acquisition Card and a ThorLabs AMP100 Transimpedance Amplifier are deployed to measure voltage and current.
The thickness of individual layers is measured with the KLA-Tencor P-7 profilometer.
The cleaning process is performed using the reactive ion etching (RIE) system from Glow Research Company.

\threesubsection{Fabrication process}\\

The substrate is cleaned for 10 minutes using sonication in isopropanol (IPA), followed by three minutes with the RIE cleaning process.
A $33~nm$ sandwich layer of Cr and Ag are thermally deposited as gate electrodes at the rate of $0.1~\AA s^{-1}$ and $1.5~\AA s^{-1}$ at the base pressure of $9\times 10^{-6}$ Torr, respectively.
The gate dielectric is obtained by the CVD deposition of Parylene diX-SR, resulting in a $400~nm$ thin film.
Organic p- and n-type semiconductors, DNTT and N1200, are deposited using the thermal evaporation process at the based pressure of $3\times 10^{-6}$ Torr at the rate of $0.09$ and $0.08~\AA s^{-1}$, respectively.
The temperature of the substrate is $60~^\circ C$ during the deposition process of DNTT.
The access to gate electrodes is obtained through the mechanical removal of the dielectric layer.
Finally, a $30~nm$ thick layer of Au is deposited as drain/source electrodes at the rate of $1.2 ~\AA s^{-1}$ at the base pressure of $9\times 10^{-6}$ Torr, followed by the thermal deposition of $30~nm$ Au tracks between the OFETs to form the LDI synaptic circuit.
The gates, sources, drains, tracks, and organic materials are patterned using shadow masking.

% %Acknowledgement

% \medskip
% \textbf{Supporting Information} \par %Please delete the Supporting Information statement if it is not applicable.
% %Please supply Supporting Information in another file.
% %Supporting information should not be provided in .tex format
\section{Supplementary information}
 \setcounter{figure}{0}

 \begin{figure}[H]
 \renewcommand{\thefigure}{S\arabic{figure}}
 \centering
 \includegraphics[height=2in]{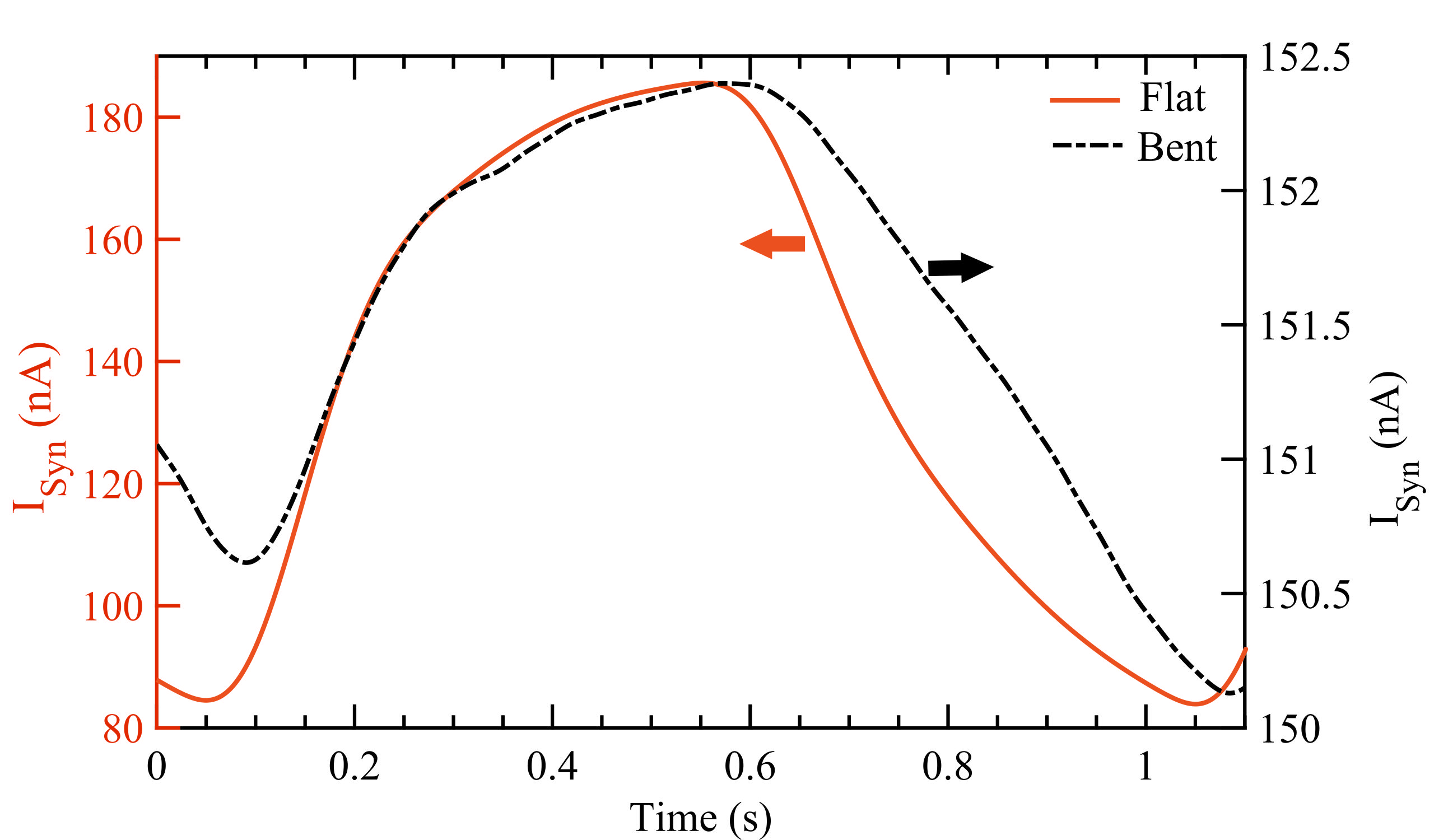}
 \caption{The synaptic current measurements ($I_{Syn}$) with $C_{Syn}=4.7~nF$, $Period_{psyn}=1~s$, and $V_W=10~V$ before (solid line) and during bending (dashed line).}
 \label{myfig12}
 \end{figure}
 \begin{figure}[H]
 \renewcommand{\thefigure}{S\arabic{figure}}
 \centering
 \includegraphics[height=2in]{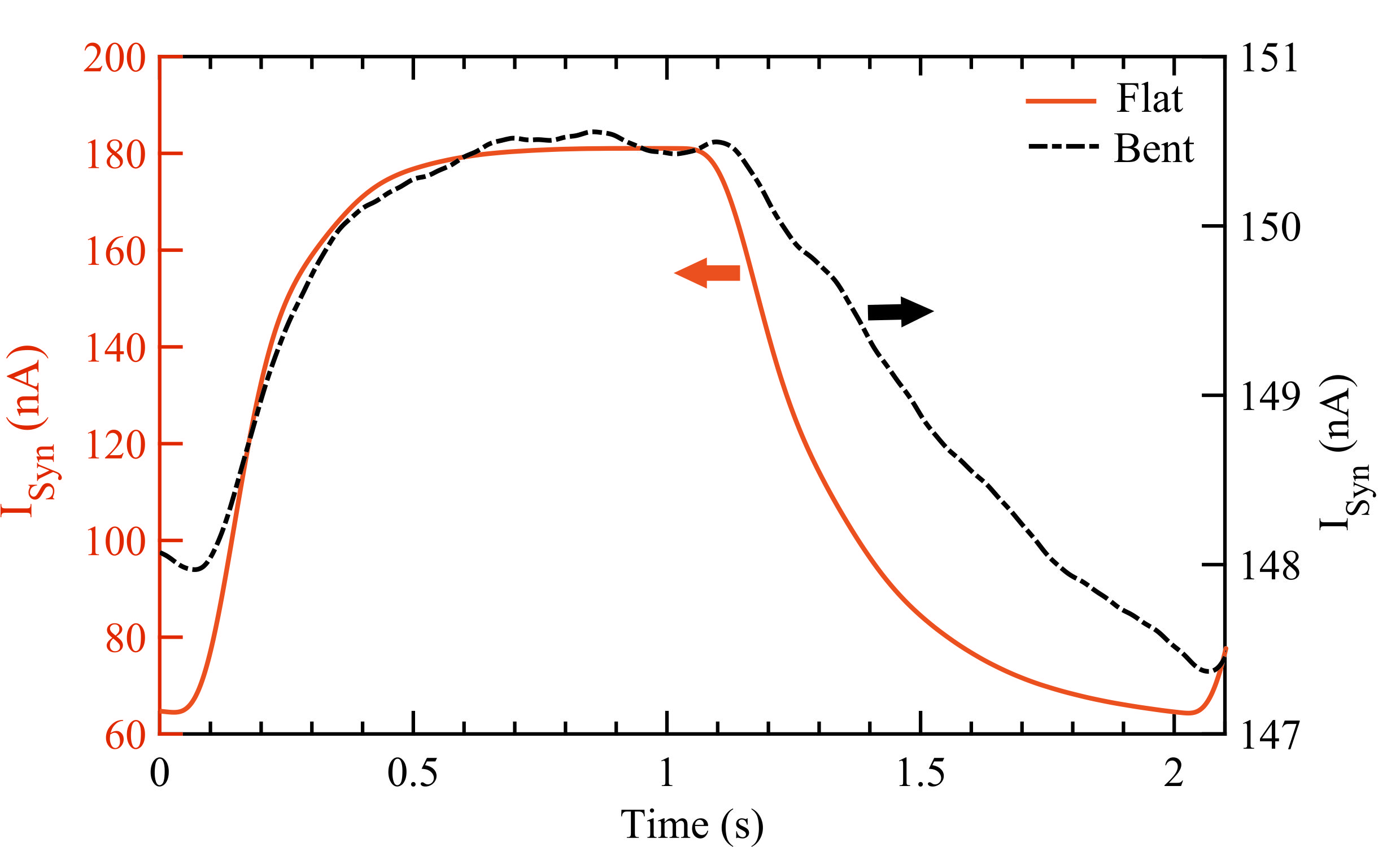}
 \caption{The synaptic current measurements ($I_{Syn}$) with $C_{Syn}=4.7~nF$, $Period_{psyn}=2~s$, and $V_W=10~V$ before (solid line) and during bending (dashed line).}
 \label{myfig13}
 \end{figure}

 \begin{figure}[H]
 \renewcommand{\thefigure}{S\arabic{figure}}
 \centering
 \includegraphics[height=2in]{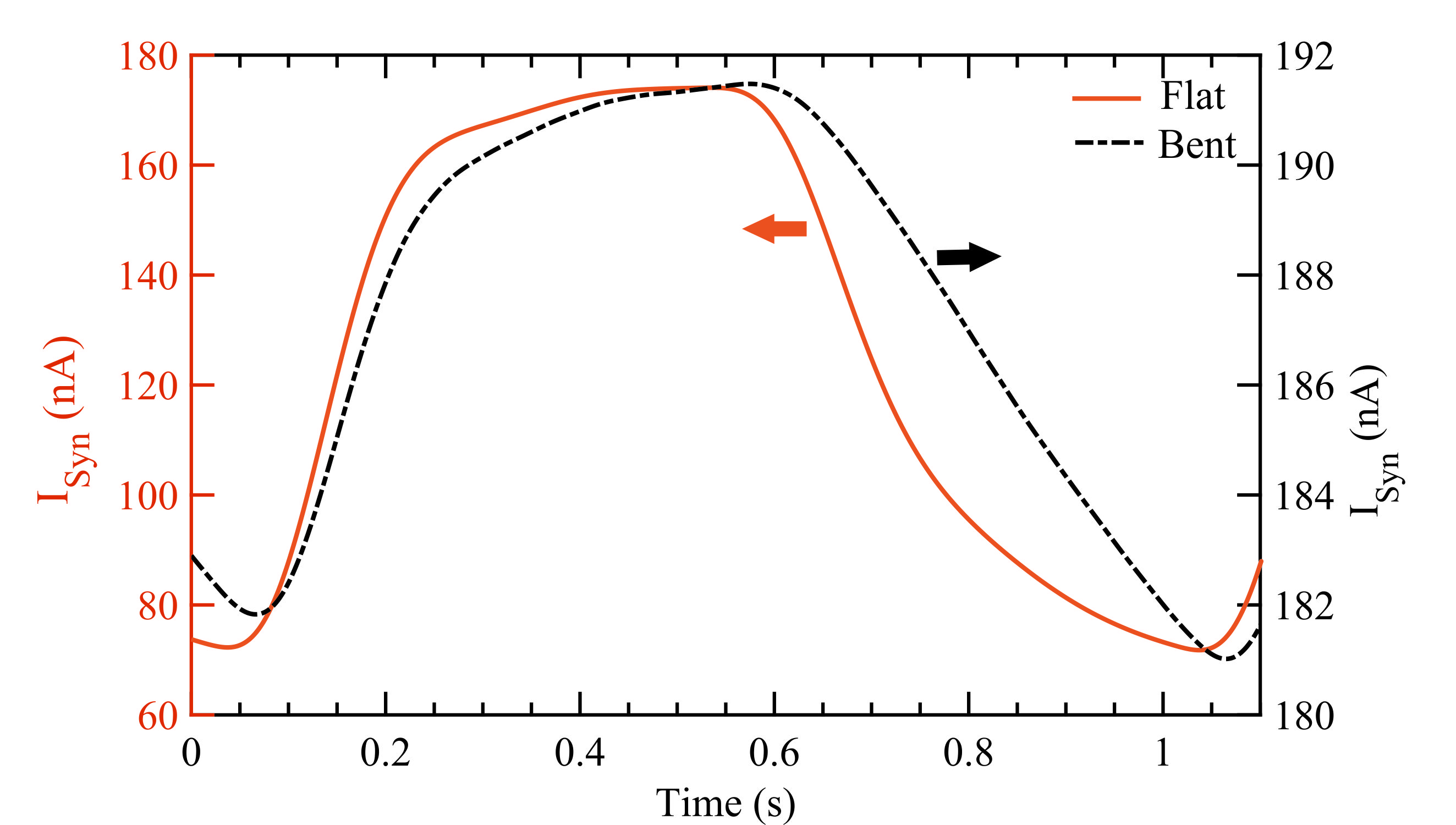}
 \caption{The synaptic current measurements ($I_{Syn}$) with $C_{Syn}=10~nF$, $Period_{psyn}=1~s$, and $V_W=10~V$ before (solid line) and during bending (dashed line).}
 \label{myfig14}
 \end{figure}

 \begin{figure}[H]
 \renewcommand{\thefigure}{S\arabic{figure}}
 \centering
 \includegraphics[height=2in]{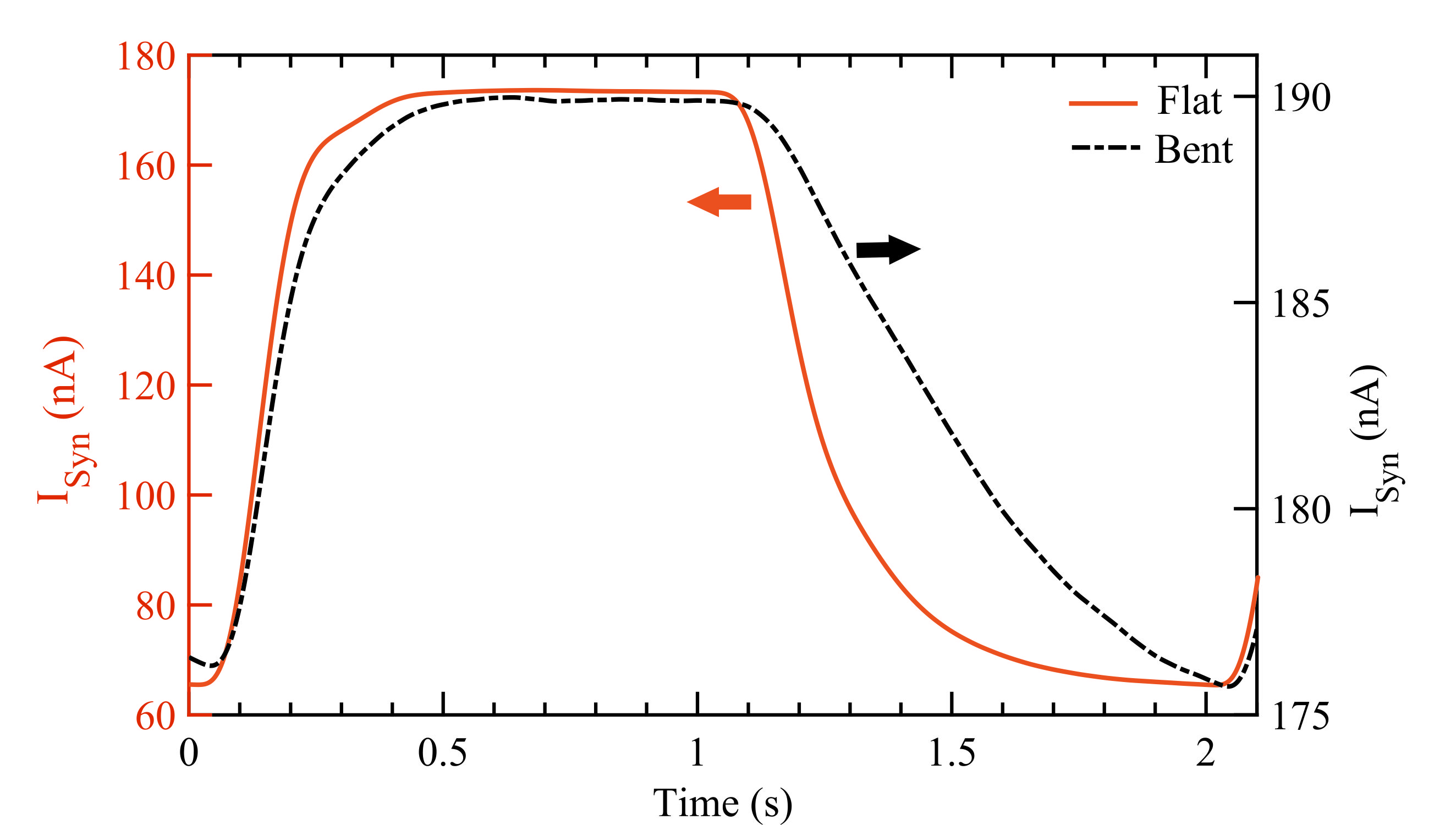}
 \caption{The synaptic current measurements ($I_{Syn}$) with $C_{Syn}=10~nF$, $Period_{psyn}=2~s$, and $V_W=10~V$ before (solid line) and during bending (dashed line).}
 \label{myfig15}
 \end{figure}

 \begin{figure}[H]
 \renewcommand{\thefigure}{S\arabic{figure}}
 \centering
 \subfloat[]{\label{fig_16a}\includegraphics[height=2in]{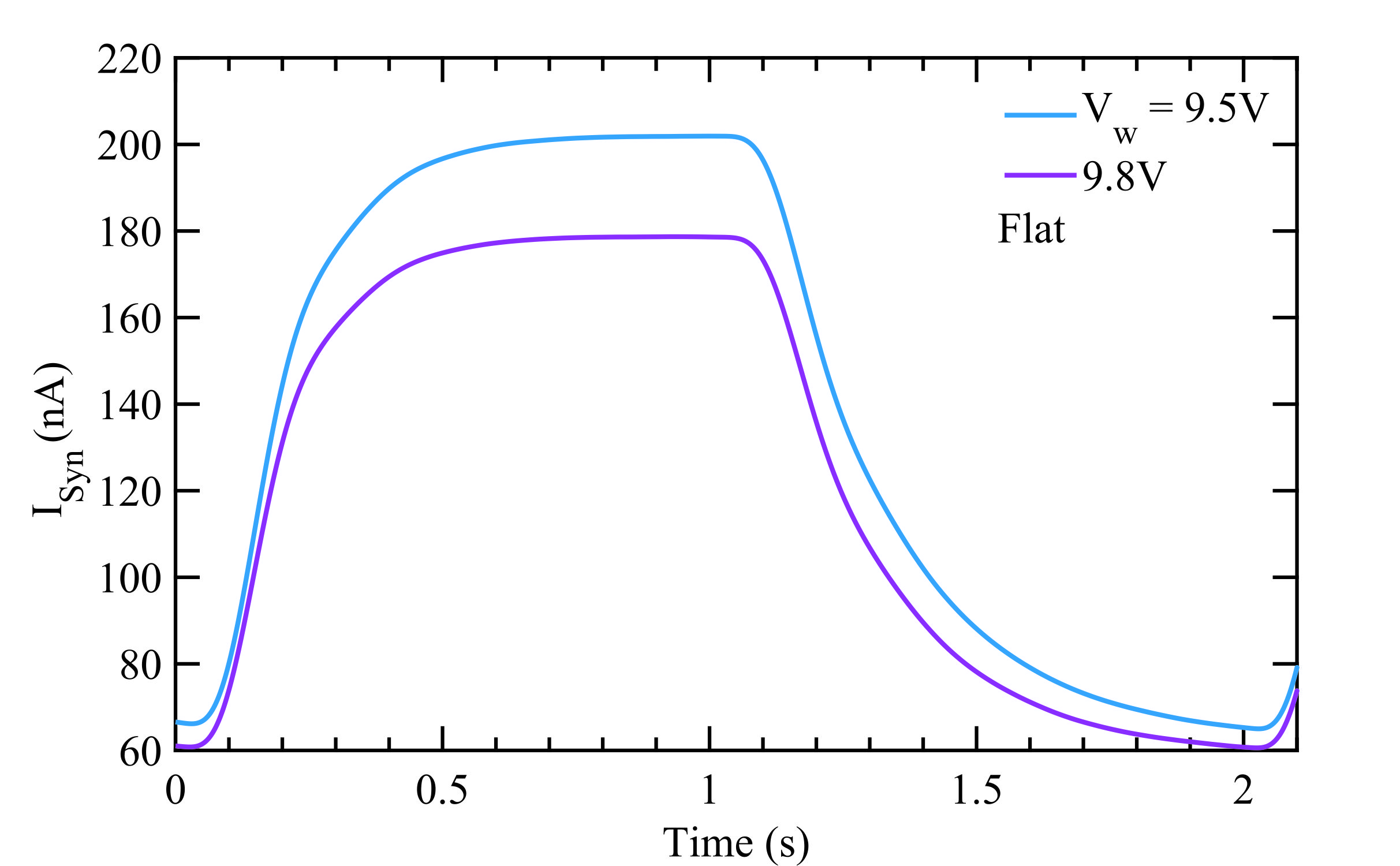}}
 \hspace{10pt}
 \subfloat[]{\label{fig_16b}\includegraphics[height=2in]{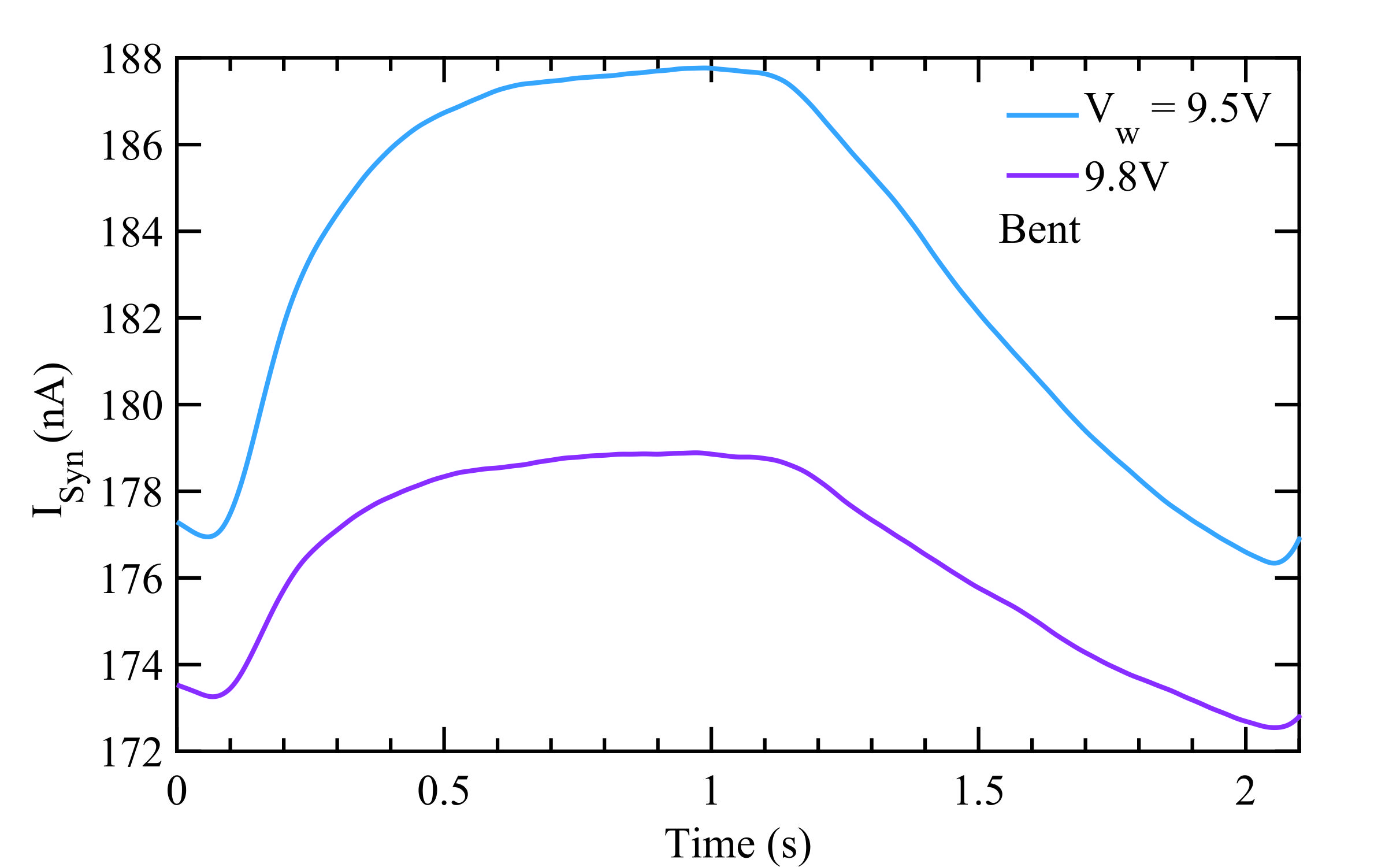}}
 \caption{The synaptic current measurements ($I_{Syn}$) with $C_{Syn}=10~nF$, $Period_{psyn}=2~s$, and $V_W=9.5~V$, and $9.8~V$.}
 \label{myfig16}
 \end{figure}

 \begin{figure}[H]
 \renewcommand{\thefigure}{S\arabic{figure}}
 \centering
 \subfloat[]{\label{fig_17a}\includegraphics[height=2in]{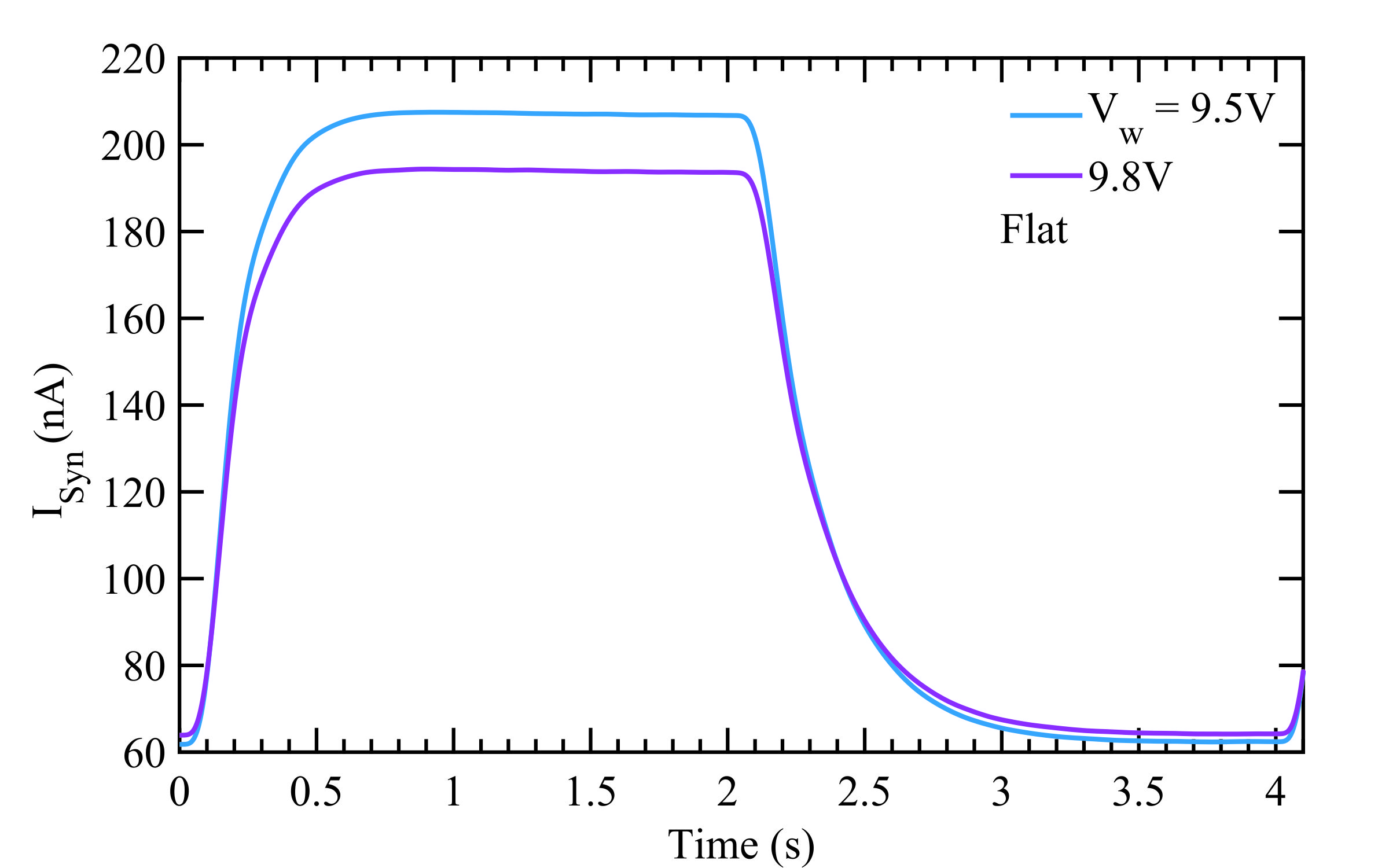}}
 \hspace{10pt}
 \subfloat[]{\label{fig_17b}\includegraphics[height=2in]{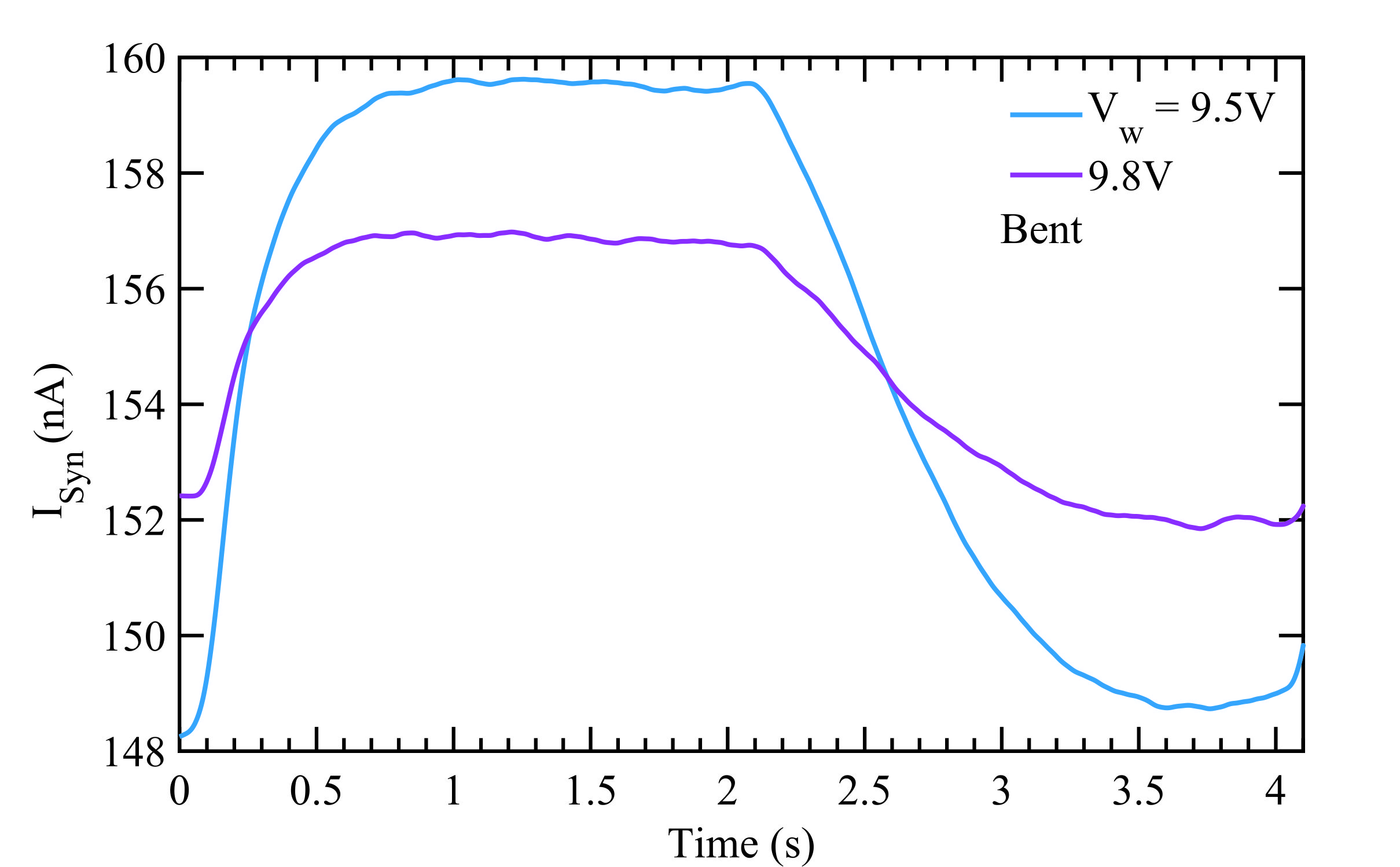}}
 \caption{The synaptic current measurements ($I_{Syn}$) with $C_{Syn}=10~nF$, $Period_{psyn}=4~s$, and $V_W=9.5~V$, and $9.8~V$.}
 \label{myfig17}
 \end{figure}

% \begin{comment}

% Nieman et al. presented a method to estimate the time constant having an interval of the step response.
% However, the method is susceptible to error \cite{b46}.
% Let's assume the step response in the time interval [$t_1$,$t_2$], and $y\:'$ is the tangent slope of a line fitted in $t_1$ to the step response.
% The time constant is estimated as follows:\
% \begin{equation} \label{eq_155}
% R_1(\Delta t) = \frac{y(t_1+\Delta t)-y(t_1)}{y\:'(t_1)\Delta t}
% \end{equation}
% \begin{equation} \label{eq_156}
% \Delta t = t_2-t_1 = \alpha \tau
% \end{equation}
% \begin{equation} \label{eq_157}
% R_1(\alpha \tau) = \frac{1-e^{-\alpha}}{\alpha}
% \end{equation}
% In \textbf{Equation~\ref{eq_156}}, $\alpha$ is a scaling factor concerning the time constant.
% Given $\alpha$ from \textbf{Equation~\ref{eq_157}} and $\Delta t$, the time constant can be extracted from \textbf{Equation~\ref{eq_158}}.
% \begin{equation}\label{eq_158}
%     \tau = \frac{\Delta t}{\alpha}
% \end{equation}

% \end{comment}

% Acknowledgements
\medskip
\textbf{Acknowledgements} \par %delete if not applicable))
This work was partially supported by the Purdue Polytechnic’s Realizing the Digital Enterprise graduate fellowship and Office of Naval Research Young Investigator Program, Award No.: N00014-21-1-2585.
The authors would also like to acknowledge the help of Prof.
Ramses Martinez, Prof.
Arman Sabbaghi, Prof.
Walter Daniel Leon-Salas and Prof.
Richard Voyles from Purdue University, for their help in device fabrication, data analysis, and fruitful discussions.

% References
\medskip
\printbibliography %Prints bibliography
% Use the following code if you wish to generate your bibliography with BibTeX;
% replace the string "MSP-template" below with the name(s) of
% the BibTeX data base(s) you want to use.
% The resulting bibliography-output (the content of the .bbl file)
% must be pasted back into this file before submission.
% Please also include your BibTeX data base file(s) in your submission
% so that we can re-run BibTeX if necessary.
%
%\bibliographystyle{MSP}
%\bibliography{ref.bbl}

% Figures/tables and captions
% Permission statements are required for all figures reproduced or adapted from previously published articles/sources.
%Please also ensure that all necessary permissions to reproduce images have been received
% Please remove these statements for original figures

% Please provide Biographies and photos for Essays, Feature Articles, Progress Reports, Reviews, and Perspectives for those authors who should be highlighted  
% These should be at most 100 words long
% For other article types this section can be removed
% Photographs should be 40mm broad and 50 mm high

\begin{figure}[h]
  \includegraphics[height=1.4in]{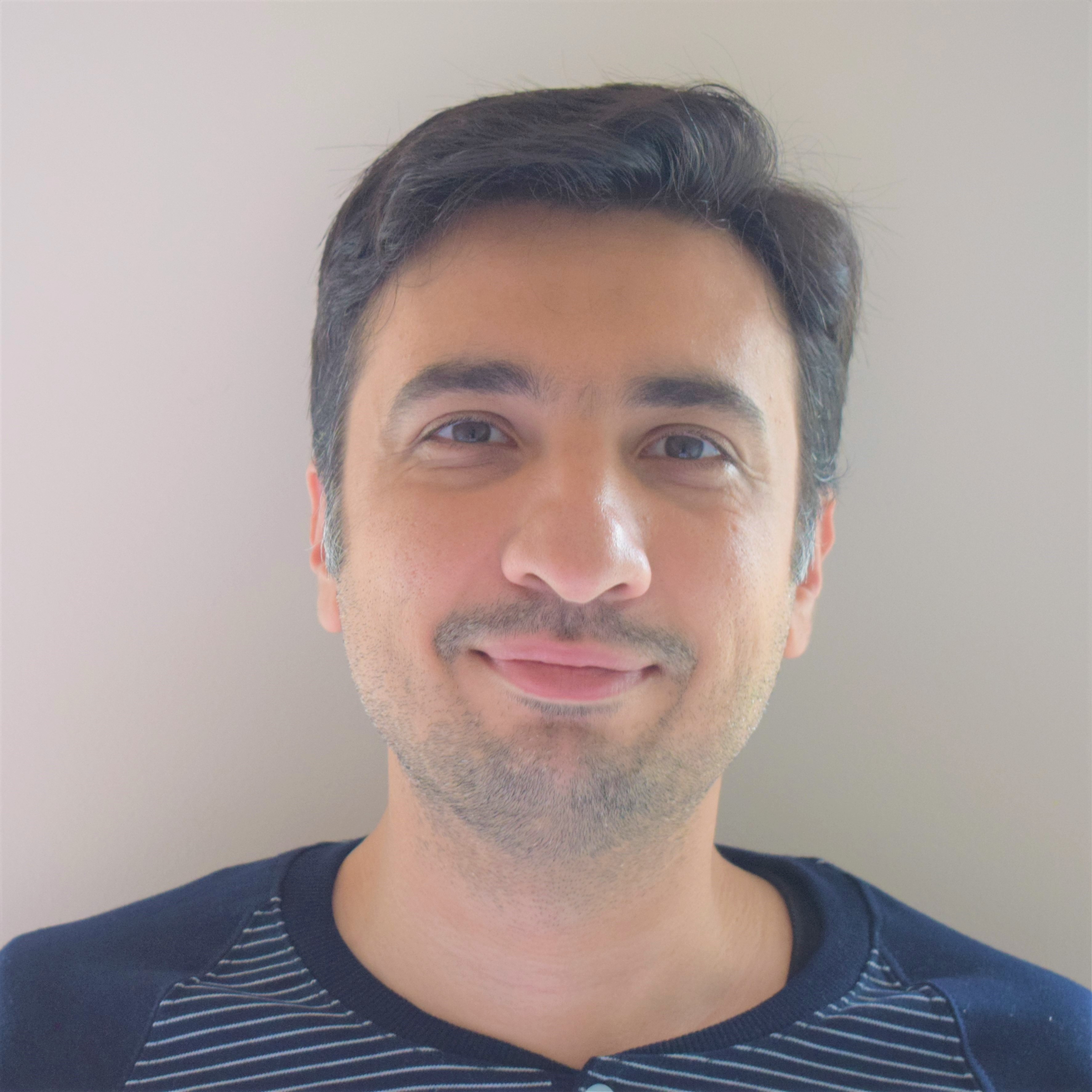}
  \caption*{Mohammad Javad Mirshojaeian Hosseini is a PhD student in the school of Engineering Technology (SoET) at Purdue University.
He obtained his Master of Science in Mechatronics from K.N.
Toosi University of Technology, School of Mechanical Engineering, Iran.
His research interests include organic brain-machine interface, organic neuromorphic electronics and its applications in neuroscience, biosensors, and flexible electronics.
He is currently developing spiking organic neuromorphic circuits and systems.
}
\end{figure}

\begin{figure}[h]
  \includegraphics[height=1.4in]{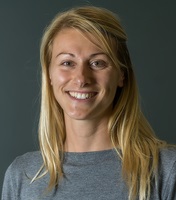}
  \caption*{Elisa Donati is a postdoctoral researcher at INI, in Zurich, Switzerland, in Neuromorphic Cognitive Systems group.
She graduated in biomedical engineering from the University of Pisa, Italy and obtained her PhD degree in Bioinspired Robotics from Institute of BioRobotics, School of Advanced Studies Pisa, Italy.
During her studies she was working on understanding how to build artificial living technologies with similar skills and computational capabilities as animals.
At INI, she is interested in the design, simulation and validation of sub-threshold VLSI circuits for biomedical applications, in particular, a biologically accurate model of respiratory CPG and ECG and sEMG signals processing and classification.}
\end{figure}

\begin{figure}
  \includegraphics[height=1.4in]{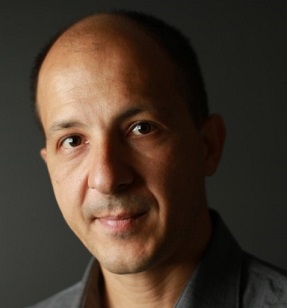}
  \caption*{Giacomo Indiveri is a Professor at the Faculty of Science at the University of Zurich, Switzerland.
He obtained an M.Sc.
degree in electrical engineering and a Ph.D.
degree in computer science from the University of Genoa, Italy.
Indiveri was a post-doctoral research fellow in the Division of Biology at Caltech and at the Institute of Neuroinformatics of the University of Zurich and ETH Zurich.
In 2006 he attained the “habilitation” in Neuromorphic Engineering at the ETH Zurich Department of Information Technology and Electrical Engineering.
He won an ERC Starting Grans on “Neuromorphic processors” in 2011 and an ERC Consolidator Grant on neuromophic cognitive agents in 2016.
His research interests lie in the study of neural computation, with particular interest in spike-based learning and selective attention mechanisms, and in the hardware implementation of real-time sensory-motor systems using analog/digital neuromorphic circuits and emerging VLSI technologies.}
\end{figure}

\begin{figure}
  \includegraphics[height=1.4in]{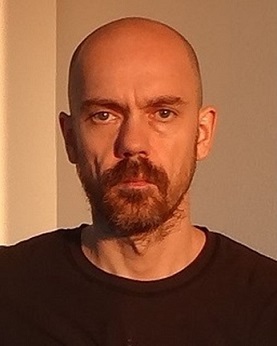}
  \caption*{Robert Nawrocki is an Assistant Professor in the School of Engineering Technology (SoET) at Purdue University.
His current research interests include physically flexible organic electronics with the application in biopotential monitoring and soft robotics, as well as neuromorphic systems, smart (meta) materialfs and neuroscience.
His research highlights include fabrication of world’s thinnest, sub 300 nm thin, organic electronics skin, fit with organic transistors and tactile sensors, and polymer neuromorphic circuitry, a biologically inspired “brain”, implemented with organic transistors and organic memristors, capable of simple pattern recognition.
He is a recipient of the 2021 Office of Naval Research, Young Investigator Award and the 2021 Ralph W.
and Grace M. Showalter Research Trust Award.}
\end{figure}
% Table of contents entry should be 50 - 60 words long
% Image should be 55 mm broad and 50 mm high or 110 mm broad and 20 mm high
%\begin{comment}

\begin{figure}
\textbf{Table of Contents}\\
\centering
\medskip
  \includegraphics[width=7in]{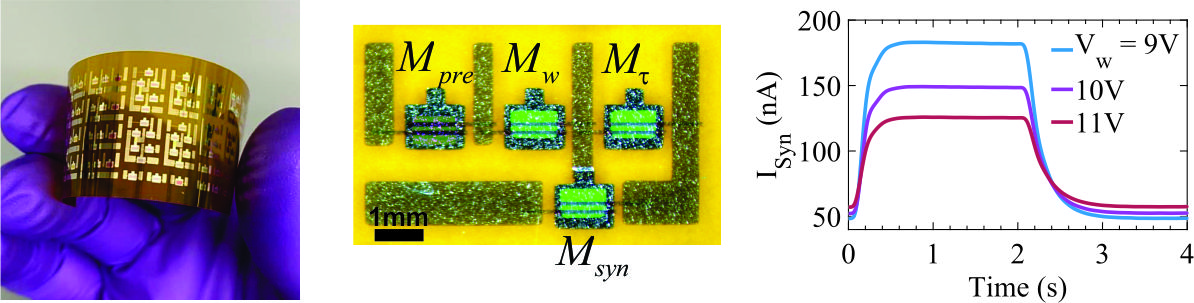}
  \medskip
  \caption*{}
\end{figure}
%\end{comment}

\end{document}